\documentclass[preprint]{aastex}

\usepackage{graphicx}

\usepackage{natbib}

\newcommand{\be}{\begin{equation}}
\newcommand{\ee}{\end{equation}}
\newcommand{\nn}{\mbox{} \nonumber \\ \mbox{} }
\newcommand{\ba}{\begin{eqnarray}}
\newcommand{\ea}{\end{eqnarray}}

\newcommand\ie{\textit{i.e.\ }}

\newcommand\lo{\mathrel{\raise.3ex\hbox{$<$}\mkern-14mu\lower0.6ex\hbox{$\sim$}}}
\newcommand\go{\mathrel{\raise.3ex\hbox{$>$}\mkern-14mu\lower0.6ex\hbox{$\sim$}}}

\begin{document}
\title{Magnetospheric Eclipses in the Binary Pulsar J0737$-$3039}
\author{M. Lyutikov}
\affil{University of British Columbia, 6224 Agricultural Road,
Vancouver, BC, V6T 1Z1, Canada}
\author{C. Thompson}
\affil{Canadian Institute for Theoretical Astrophysics,
University of Toronto, Toronto, ON, Canada}

\begin{abstract}
In the binary radio pulsar system J0737$-$3039, the faster pulsar A is 
eclipsed once per orbit.  A clear modulation of these eclipses at the 2.77 s 
period of pulsar B has recently been discovered. We construct a simple 
geometric model which successfully reproduces the eclipse light curves, based 
on the idea that the radio pulses are attenuated by synchrotron 
absorption on the 
closed magnetic field lines of pulsar B. The model explains most of the 
properties of the eclipse:  its asymmetric form, the nearly 
frequency-independent duration, 
and the modulation of the brightness of pulsar A 
at both once and twice the rotation frequency of pulsar B in different parts 
of the eclipse.
This detailed agreement confirms the 
 dipolar structure of the star's poloidal magnetic field.
 The inferred  parameters are: inclination angle between the line of
 sight and orbital
 plane normal $\sim 90.5^\circ$; inclination of pulsar B rotation axis to the
orbital plane normal $\sim 60^\circ$; and
   angle between the  rotation axis and magnetic moment  $\sim 75^\circ$.
The model makes clear predictions for the degree of 
linear polarization of the transmitted radiation.

The weak frequency dependence of the eclipse duration implies that the
absorbing plasma is  relativistic, with a density much larger than the 
corotation charge density.  Such hot, dense plasma can be effectively stored
in the outer magnetosphere, where cyclotron cooling is slow.  
The gradual loss of particles inward through the cooling radius
is compensated by an upward flux driven by a fluctuating component
of the current, and
by the pumping of magnetic helicity on the closed field lines.
The trapped particles are heated to relativistic
energies by the damping of magnetospheric turbulence and, at a slower
rate, by the absorption of the radio emission of the companion pulsar.  
A heating mechanism is outlined which
combines electrostatic acceleration along the magnetic field with the
emission and absorption of wiggler radiation by charged particle bunches.
\end{abstract}

\date{}

\section{Introduction}

The double pulsar system PSR J0737$-$3039A/B contains a recycled 22.7 
ms  pulsar (A) in a 2.4 hr orbit around a 2.77 s pulsar (B)  
\citep{burgay03,lyne}.  Pulsar A is eclipsed once per orbit, 
for a duration of $\sim 30$ s 
centered around superior conjunction.
The width of the eclipse is a weak function of the observing frequency
\citep{kaspi}. Recently, the pulsar A radio flux has been found to be
modulated by the rotation of pulsar B during the eclipse \citep{mcla04}.
The eclipse is longer when the magnetic axis of
pulsar B is approximately aligned with the 
line of sight (assuming that radio pulses are generated near the
magnetic axis of B). In addition,
there are  narrow, transparent windows in  which 
the flux from pulsar A rises nearly to the unabsorbed level.
These spikes in the radio flux are tied to the
rotation of pulsar B, and provide key constraints on the geometry of the
absorbing plasma.

The physical width of the region which causes this periodic modulation is
comparable to, or smaller, than the estimated diameter of the magnetosphere
of pulsar B.  Combined with the rotational modulation, 
this provides a strong hint that the absorption is occurring
{\it within} the magnetosphere of pulsar B, which retains
enough plasma to effect an eclipse.  An alternative 
possibility, which we disfavor, is that the eclipse is caused by 
absorption in the relativistic wind 
of pulsar A after it is shocked and decelerated by its interaction 
with the magnetosphere of B \citep{lyut,arons}.

Distinguishing
between these two models is greatly aided by a quantitative 
model of the
eclipse light curve.  In this paper, we construct such a model
and show that the light curve is consistent, in considerable detail, with 
synchrotron absorption in the pulsar B magnetosphere.
The agreement is detailed enough to constitute direct evidence 
for the presence of a {\it dipolar} magnetic field around pulsar B.   
We assume that the 
absorbing plasma is confined within a set of poloidal field lines
that is rotationally symmetric about the magnetic dipole axis.  
Considerations of the sourcing and heating of this plasma suggest
that its density will exceed the Goldreich-Julian density 
 by a few 
orders of magnitude, and that the absorbing particles will be
relativistic.  Similar conclusions has been reached independently 
by \cite{rafikov} under different assumptions about the sourcing and heating mechanisms.
Note that if the plasma has a finite transverse temperature, then
particles will be reflected from regions of high magnetic field and
trapped on the closed field lines.  Since the trapped particles cool slowly,
a large equilibrium density can be established.

From this perspective, it is clear that the
eclipse model of \cite{lyut} and \cite{arons} had to 
make three fine-tuning assumptions.  First the wind of pulsar A must be
extremely dense:  about $\sim 10^6$ times the Goldreich-Julian density
at the pulsar speed of light cylinder, in contrast with models of pair creation
which suggest values closer to $\sim 10^3$ \cite[e.g.][]{harding,hibs01}.
Second, the pulsar A wind must be slow, with typical Lorentz factor 
$\gamma_{\rm A}\sim 10$, while at the distance of $\sim 1000$ 
light cylinder radii of pulsar A it is expected that 
$ \gamma_{\rm A}\sim 1000$ \citep{mich69}.  Third, the orbital 
inclination angle is $\leq 86^\circ$;  new data indicate that  the 
inclination angle is very close to $90^\circ$ \citep{coles,ransom04}.
The model does not offer a simple explanation for the observation
transparent windows in the eclipse light curves: while the position
of the magnetosheath is expected to vary by $\sim  25 \%$ between 
different phases, it is difficult to see how the the sheath can become 
fully transparent.   In addition, the sheath is expected to be wider when 
the magnetic axis is perpendicular to the line of sight while, in contrast 
to this, the eclipse is narrow at such points \citep{mcla04}.

The presence of a such a dense plasma within the magnetosphere of pulsar
B is contrary to what is usually claimed for isolated radio pulsars. 
Thus, a source of charged particles --- of both signs --- is required
on the most extended closed magnetic field lines.
We describe two related mechanisms by which the absorbing
particles can be supplied by the star itself, or by pair cascades 
close to its surface.\footnote{It is unlikely that closed field lines 
are seeded by particles from  pulsar A wind without, at the
same time, producing unwanted absorption in the much wider
region in between the magnetopause and the wind shock \citep{lyut}.}
First, torsional deformations
of the pulsar B magnetosphere, driven by turbulence in the surrounding
magnetosheath, will generate a fluctuating
component of the current, which modulates the effective opening
angle of the field lines close to the magnetic poles, and the 
Goldreich-Julian current.
A second, more subtle, effect involves the pumping of magnetic
helicity  from the outer to the inner magnetosphere.
We outline a simple energetic reason why this will occur
when the outgoing and return currents are not precisely balanced
in each magnetic hemisphere.   The dipolar magnetic field then supports
a static twist, and a persistent current flows along magnetic field lines
which close within the magnetosphere.


The plasma on closed field lines is quickly heated to relativistic energies.
The outer magnetosphere
of pulsar B is heated both by the damping of relativistic turbulence,
and by the resonant absorption of the radio waves from pulsar A.
(The resonant absorption of high energy emission from pulsar A
can occur only very close to the neutron star, where the geometric
cross section is small.)  Since the energy flux in the wind of
pulsar A exceeds the energy flux in its radio emission by several
orders of magnitude, we argue that turbulent heating is likely to 
dominate -- especially in regions of high synchrotron optical depth.

Turbulence in the surrounding magnetosheath generates Alfv\'en waves
 on closed field lines. The non-linear interaction between the waves 
creates higher-wavenumber modes and, eventually, the dissipation of 
turbulent energy on small scales.  Inside the magnetopause, where
the pressure of the background magnetic field exceeds the particle
pressure, the inner scale of the turbulent
spectrum occurs when the current fluctuations become too large to be 
supported by the available free charges: the plasma becomes charge-starved
at small scales \citep{thomp98}.  As a result, charges are
accelerated electrostatically along the background magnetic field.
Bunches of such relativistic charges will generate high-brightness
radio emission (wiggler radiation) when they scatter off the highest
frequency Alfv\'en modes.  We show that this radiation can self-consistently
excite the transverse motion of the particles.  The conclusion is that
self-absorbed radio photons that are generated {\it within}
the optically thick plasma are the dominant heating mechanism.

The plan of the paper is as follows.  In \S \ref{basic}
we review the basic parameters of the PSR J0737$-$3039 system.
In the following section, we describe how electric currents may
be excited in the magnetosphere of a neutron stars, and
how the particles which support the current are supplied.
 In \S \ref{reabs} we describe particle heating at large 
synchrotron optical depth, and calculate the distribution of 
synchrotron-absorbing particles that is used to model the eclipse.
Section \ref{geom} presents our calculations of the eclipse light curves.
In \S \ref{ions} we discuss  modification of the model due to presence
of ions in magnetosphere.
The polarization of the radio waves transmitted through the
absorbing magnetosphere is calculated in \S \ref{Polariz}.  
In \S \ref{disc} we discuss  predictions  and compare with 
alternative eclipse models.
In the final section \ref{concl} we
 summarize 
the broader implications of our work.

\section{Basic Parameters}\label{basic}

The semi-major axis of the J0737$-$3039 system is 
$a \sim 8.8\times 10^{10}$ cm, the orbital period is $2.45$ hr, and the 
orbital eccentricity is small, $e \simeq 0.088$.  At the time of eclipse, 
the separation of the two pulsars is $D_{AB} \simeq 8\times 10^{10}$ cm, 
their relative orbital velocity is $V_\perp = 680$ km s$^{-1}$ transverse 
to the line of sight, and one orbital degree near eclipse corresponds to 
a time interval of $20.5$ seconds.
The observed eclipses cover $\Delta\phi \simeq 0.035$ 
radians (2 degrees) of orbital phase.
The corresponding half-width of the eclipsing material, transverse
to the line of sight,  is $({1\over 2}\Delta\phi) D_{AB} = 1.4\times 10^9$ cm.
This is much smaller than the pulsar B light cylinder, which has
a radius $c/\Omega_B = 1.3 \times 10^{10}$ cm.
The orbital inclination is close to $90^\circ$ \citep{coles,ransom04}.

The spin periods and period derivatives, $P_{A,B}$ and 
$\dot P_{A,B}$, have been measured for the two pulsars.
Normalizing the moments of inertia $I_{A,B}$ to $10^{45}$ g cm$^2$,
the spindown luminosities are
$L_A  = I_A \Omega_A \dot{\Omega}_A =  
5.8 \times 10^{33}\,I_{A,45}$ ergs s$^{-1}$ and
$L_B = I_B \Omega_B \dot{\Omega}_B  = 1.6 \times 10^{30}\,I_{B,45}$
ergs s$^{-1}$ \citep{burgay03}.

The magnetosphere of pulsar B is  truncated, compared with  the 
magnetosphere of an isolated pulsar of the same spin,
by the relativistic wind flowing outward from pulsar A.
The spindown torque of pulsar B is therefore modified.
The surface magnetic field can
be estimated self-consistently only by modeling the interaction
between wind and magnetosphere.
Different  effects can contribute 
to the actual torque \citep{lyut,arons}.  
For example, the open-field current is increased compared with that of 
an isolated dipole, which implies
\be
L_B
= {\cal N}_B  B_{\rm NS}^2 R_{\rm NS}^2  c 
\left( { \Omega_B R_{\rm NS} \over c}\right)^2
\left( { R_{\rm NS} \over R_{\rm mag} } \right)^2.
\label{L2}
\ee
Here $B_{\rm NS}$ is the average surface dipole field, and
${\cal N}_B$ is a parameter that rescales the torque acting on pulsar B.

The magnetospheric radius $R_{\rm mag}$ is determined by the pressure balance
between the wind of pulsar A and the magnetic pressure of pulsar B 
evaluated at,
\be
{B^2_{\rm mag} \over 8 \pi} =
{I_A \Omega_A \dot{\Omega}_A \over 4 \pi c D_{AB}^2}
\label{Bm}
\ee
Thus, magnetic field at $R_{\rm mag}$ depends only on parameters of pulsar A wind. 
Using our fiducial numbers we find
\be
B_{\rm mag}= \sqrt{2 I_A \Omega_A \dot{\Omega}_A \over  D_{AB}^2 c} =  7\,
I_{A,45}^{1/2} \;\;\;\; {\rm G}.
\ee
On the other hand, magnetospheric radius $R_{\rm mag}$  and surface magnetic field  $B_{\rm NS}$
depend on  details of the wind-magnetosphere interaction. Using Eqs.
(\ref{L2}-\ref{Bm}) we find
\ba &&
B_{\rm NS}  
\simeq {D_{AB}^{1/2} \left( I_B  \dot{\Omega}_B/\Omega_B \right)^{3/4} c
     \over {\cal N}_B^{3/4} \left( I_A \Omega_A \dot{\Omega}_A \right)^{1/4} R_{\rm NS}^3  }
 = 4 \times 10^{11} {\cal N}_B^{-3/4}\,{I_{B,45}^{3/4}\over I_{A,45}^{1/4}}
    \;\;\;\; {\rm G};
\nn &&
 R_{\rm mag} \simeq  \left[ { (cD_{AB})^2 I_B  \dot{\Omega}_B \over 
{\cal N}_B I_A \Omega_A \dot{\Omega}_A \Omega_B} \right]^{1/4}
= 4.0 \times 10^9\, {\cal N}_B^{-1/4}\,
\left({I_B\over I_A}\right)^{1/4}\;\;\;\; {\rm cm}
\label{Bn}
\ea
Below we will use these parameters as our fiducial numbers.


\section{Injection of Plasma on Closed Field Lines}
\label{EM0}

To calculate the eclipse light curve, one needs the basic plasma
properties in the outer magnetosphere of pulsar B -- most importantly,
the equilibrium density, particle energy distribution, and composition.
In this and the next section,
we motivate our basic assumption that this confined plasma
has a large synchrotron
optical depth and is relativistically hot.  The implications of
a significant ion component of the plasma are addressed in \S \ref{ions}.

One advantage of positioning
the eclipsing plasma on the closed magnetic field lines of pulsar B
is that a large
particle density can build up slowly, over many rotation periods.
Heating of the particles in the outer magnetosphere, where cooling
is long, allows them to be trapped through the effect of magnetic bottling.
Qualitatively, if particles are leaking out of the
outer parts of  magnetosphere on timescale $\lambda$ times the
pulsar period, then a source of particles that generates the Goldreich-Julian 
density each period would result in an equilibrium multiplicity
$\lambda$.  The typical life times of particles in the outer 
magnetosphere are indeed long, $\sim 10^5$ spin periods, see eq. (\ref{p}).
Since the required density at the edge of the magnetosphere is 
$\sim 10^5$ times the Goldreich-Julian density (see \S \ref{succ}),
plasma should be sourced at a rate that is comparable with the  
Goldreich-Julian density per period, or higher.
This sourcing can be driven either by torsional Alfv\'en waves
(AC sourcing); or by a pumping of magnetic helicity 
from the outer to the inner closed field lines (DC sourcing).

\subsection{Electrodynamics of a Perturbed Magnetosphere}
\label{EM}

The magnetopause of pulsar B
is expected to be strongly turbulent, as particle-in-cell simulations
demonstrate (Arons, Spitkovsky, \& Backer 2004, in preparation). 
Turbulence in the sheath will launch (torsional) Alfv\'en
waves on the closed magnetic field lines 
which carry charge and current.  The sign of the current alternates
on a timescale comparable to the spin period $P$, which is a
few times longer than $\sim R_{\rm mag}/c$.  It will also alternate
on a timescale comparable to the flow time behind the wind
shock (which is comparable to or somewhat larger than $P$).
The depth to which turbulence may be excited in the magnetosphere
of pulsar B will depend on more subtle details, such as the
resonant interaction between compressive and torsional modes.
We will not examine them in this paper.

Torsional Alfv\'en waves, launched  by turbulence in the magnetosheath,
impose a twist on closed field lines. 
If the typical fluctuating component of the magnetic field in the wave at 
radius $R_{\rm mag}$ is $B_\phi = \Delta\phi B_P$,
then there is an associated current
 $J \sim c B_\phi / 4\pi R_{\rm mag} = \Delta\phi c B_P  / 4\pi R_{\rm mag}$.
(Here $B_\phi$ is the toroidal magnetic field, and $\Delta\phi$ is the angle
through which the poloidal magnetic field $B_P$ is twisted.)
As the twist in the magnetic field (the Alfv\'en wave) propagates 
along the closed field line toward the star, the associated
current density increases.  The ratio of the 
current density to the local corotation current density remains 
approximately constant,
\be
{J\over \rho_{\rm GJ}c} \simeq {\Delta\phi\over\cos\theta_B}\,
\left({\Omega_B R_{\rm mag}\over c}\right)^{-1}
= 1.1\,{\cal N}_B^{1/4}\,\left({\Delta\phi\over 0.1}\right) \,
\left({\cos \theta_B \over 0.3}\right)^{-1}
\label{J}
\ee
Here
\be\label{corot}
\rho_{\rm GJ} = -{{\bf\Omega}\cdot{\bf B}\over 2\pi c}
 = - \cos \theta_B {\Omega_B B \over 2\pi c}  
 \label{nGJ}
\ee
is the corotation charge density \citep{goldreich69}, and
$\theta_B$ is angle between the local magnetic field and ${\bf\Omega}$.  
The net current is
\be
I \sim {\Delta\phi\over 2} cB_{\rm NS}{R_{\rm NS}^3\over R_{\rm mag}^2}.
\ee

In order to supply this current,
a minimum particle  density is required.  If at some point 
the actual particle density is below the minimum value $|\vec\nabla\times
\vec B|/4\pi e$, then
charges will be supplied from the star or created locally through
the reaction\footnote{Because pulsar B is older than $\sim 10^8$ yrs, 
and its spindown power is small,
the density of target X-ray photons is probably too small to allow 
pair creation via the channel $\gamma + \gamma \rightarrow e^+ + e^-$} 
$\gamma \rightarrow e^+ + e^-$.   
A minimum charge density is always present on closed field lines,
the corotation  charge density  (\ref{nGJ}). 
When pulsar B is nearly an orthogonal rotator (as is implied by our
eclipse model; see \S \ref{geom}), $\theta_B$ is close to 90 degrees,
and the corotation charge density can fail to supply
the fluctuating current even for a small twist angle $\Delta\phi 
\sim 0.1$.  In this case, the current can not be supplied by
an upward or downward drift of the corotation charge;  instead
charges of both signs will be supplied to the outer magnetosphere.
The sign of the instantaneous
charge flow will fluctuate in tandem with the sign of the current.

Electrons and positrons cool slowly by cyclo/synchrotron
emission in the outer parts of the magnetosphere.  Beyond the cooling 
radius,
\be\label{rcool}
R_{\rm cool} =  1.1\times 10^8 \,
\left[ {(B_{\rm NS} R_{\rm NS}^3)_B \over 5 \times 10^{29}~{\rm G~cm^3}}
\right]^{2/5}\, {\rm cm},
\label{cool}
\ee
the cyclotron timescale $t_{\rm cyc} =
3m_e^3 c^5 / 2e^4 B^2$ becomes shorter than $\sim r/c$.
Only those particles reaching the
cooling radius can precipitate to the surface.  If the particles
in the outer magnetosphere are isotropized, then the
cooling fraction is small (\S \ref{bottle}),
\be
f_{\rm cool} \sim {1\over 8}\left({R_{\rm cool}\over R_{\rm mag}}\right)^3
= 2.6\times 10^{-6} \left[{(B_{\rm NS}R_{\rm NS}^3)_B
\over 5 \times 10^{29}~{\rm G~cm^3}}\right]^{6/5}\,
\left({R_{\rm mag}\over 4\times 10^9~{\rm cm}}\right)^{-3}.
\label{p}
\ee
The upward flux of particles is shut off
when the local density at  $R_{\rm cool}$ exceeds the density
$|J|/ec = \lambda_J n_{\rm GJ}(R_{\rm cool})$ that is needed to
supply the entire current close to the neutron star.  On this basis,
we obtain an estimate of the equilibrium multiplicity of electrons
(or positrons)
\be\label{multa}
\lambda_{\rm eq}(r) = \lambda_J {B(R_{\rm cool})\over B(r)}
= 9.6\times 10^4\,\lambda_J \, \left[ { (B_{\rm NS}R_{\rm NS}^3)_B
             \over 5 \times 10^{29}~{\rm G~cm^3}} \right]^{-6/5}\,
\left({r\over 4\times 10^9~{\rm cm}}\right)^3,
\label{lambdamax}
\ee
in the slow-cooling region.
Here $r$ is the distance from pulsar B at the magnetic equator.
Notice that this expression is independent of the electron
temperature in the outer magnetosphere.
It also implies a characteristic number density,
\be
n_e = \lambda_{\rm eq}\,\cos\theta_B\,{\Omega_B B\over 2\pi ec}
     = 1.9\times 10^4\;\lambda_J\,\cos\theta_B\,
        \left[ { (B_{\rm NS}R_{\rm NS}^3)_B
             \over 5 \times 10^{29}~{\rm G~cm^3}} \right]^{-1/5}\;\;\;\;
{\rm cm^{-3}},
\label{ne}
\ee
that is approximately independent of $r > R_{\rm cool}$.

An independent upper bound on the multiplicity comes 
from the requirement that the particle pressure not exceed the magnetic
dipole pressure:
\be\label{multc}
{\lambda}_{\rm max} = {eB(r)c\over (4\cos\theta_B) k_{\rm B}T_e \Omega_B} 
=
{1.5\times 10^7\over\cos\theta_B}\,\left[ { (B_{\rm NS} R_{\rm NS}^3)_B
\over 5 \times 10^{29}~{\rm G~cm^3}} \right]\,
\left({k_{\rm B}T_e\over m_ec^2}\right)^{-1}
\left({r\over 4\times 10^9~{\rm cm}}\right)^{-3}.
\ee
Electrons/positrons with a density ${\lambda}_{\rm eq}$ have, 
equivalently, a maximum temperature
\be\label{ktmax}
{k_{\rm B}T_{e,\rm max}\over m_ec^2} =
{15\over\lambda_J\cos\theta_B} \left[ { (B_{\rm NS} R_{\rm NS}^3)_B
\over 5 \times 10^{29}~{\rm G~cm^3}} \right]^{11/5}
\left({r\over 4\times 10^9~{\rm cm}}\right)^{-6}.
\ee
 
The supply of charges to the magnetosphere can be greatly augmented
by multiple stages of pair creation \citep[e.g.][]{hibs01}.
This requires the formation of large gaps
(regions of uncompensated parallel electric field) in the
inner magnetosphere.  
From the perspective of the near-field electrodynamics of the
neutron star, the twist in the magnetic field 
is modulated only very slowly (over
$\sim 10^4$ times the light-crossing time of the star).
This part of the circuit will therefore bear some resemblance
to the  polar cap of an ordinary, isolated radio pulsar.  

Electric gaps can appear higher in the magnetosphere, if the particles 
at a given point have a fixed sign and
a charge density equal to $\rho_{\rm GJ}$. 
These gaps
will form along the surfaces where ${\bf \Omega}\cdot{\bf B} = 0$
 \citep[e.g.][]{Holloway77}.
The magnitude of the electrostatic potential drop that builds
up across this gap during one torsional oscillation would 
approach $\Delta\Phi \sim eBR \Delta\phi$.  However, the presence of
a dense, polarizable plasma would largely prevent these gaps
from forming on the closed magnetic field lines,
outside the cyclotron cooling radius (\ref{rcool}).
The gaps would also be suppressed if the magnetosphere supported
a {\it persistent} twist with a magnitude greater than
$\Delta\phi \sim \Omega r/c$, a possibility that we
entertain in Appendix \ref{Hpump}.

\subsection{Comparison: Absorption of Particles from the Wind of pulsar A}

It is also worth considering the wind of pulsar A as a competing source
of absorbing particles in the magnetosphere of pulsar B.  Transfer
of particles from wind to magnetosphere will occur rapidly through
reconnection between the oscillating magnetic field that is advected
by the wind, and the forward surface of the magnetosphere 
\citep[e.g.][]{thomp94}.
The particle density in the pulsar A wind at the position of pulsar B
is 
\be
n_A
= { {\lambda}_A \over D_{AB}^2}  \sqrt{  I_A \Omega_A \dot{\Omega}_A   \over 4 \pi c e^2} 
= 4\times 10^{-2} \lambda_A {\rm cm^{-3}}.
\ee
Here ${\lambda}_A$ is the multiplicity
of pairs created by a cascade on the open field lines of pulsar A.
Comparing with the characteristic charge density 
$n_B = \lambda_{\rm eq} n_{{\rm GJ},B} = 
\lambda_{\rm eq}\Omega_B B(R_{\rm mag})/2\pi ec$ 
in the magnetosphere of pulsar B, at a distance $R_{\rm mag}$ 
from the star, one has
\be
{n_B(R_{\rm mag})\over n_A}
\sim  5\;{{\lambda}_{\rm eq}\over{\lambda}_A}\,
\left({R_{\rm mag}\over 4\times 10^9~{\rm cm}}\right)^{-3}\,
\left[{(B_{\rm NS} R_{\rm NS}^3)_B\over 5\times 10^{29}~{\rm G~cm^3}}\right].
\ee
The density of particles trapped from the wind of pulsar A will
not, generally, exceed the density in the wind itself:  otherwise
the return of particles from the magnetosphere back to the wind
will balance the gain.
One expects that ${\lambda}_{\rm eq} > 10^5 > {\lambda}_A$, and so
we conclude that the dominant source of absorbing particles in the
magnetosphere of pulsar B is likely to be the particles pulled
outward from its inner magnetosphere.


\section{Particle Heating in the Outer Magnetosphere}
\label{reabs}


There are two principal sources of free energy in the
outer magnetosphere of pulsar B: the relativistic wind emitted
by pulsar A, and the radio emission of pulsar A.
Both can be effective at heating trapped particles 
to relativistic energies.  The wind energy flux is, nonetheless
several orders of magnitude larger (in spite of the uncertain effects
of wind collimation and beaming).  In addition, the equilibrium
plasma density estimated in \S \ref{EM0} is too large to
allow effective heating through radio absorption throughout the
bulk of the pulsar B magnetosphere.

The kinetic power incident on the magnetosphere
(radius $R_{\rm mag}$; eq. \ref{Bn}) at a separation 
$D_{AB} = 8\times 10^{10}$ cm is
\be
P_{\rm wind} \;\simeq\; \left({R_{\rm mag}\over 2 D_{AB}}\right)^2\,L_A
\;=\; 4\times 10^{30}\,{\cal N}_B^{-1/2}\,(I_{A,45}\,I_{B,45})^{1/2}
\;\;\;\;{\rm ergs~s^{-1}}.
\ee
 The radio energy flux incident on pulsar B
is $\nu F_\nu  = (D_{\rm J0737}/D_{AB})^2 \,(\nu F_\nu)_{\rm obs} 
\simeq 1.2\times 10^4\,(D_{\rm J0737}/0.6~{\rm kpc})^2$ ergs cm$^{-2}$
s$^{-1}$ at 1.4 GHz, where $(F_\nu)_{\rm obs} \sim 1.6$ mJy  is the
observed flux at this frequency, and $D_{\rm J0737} \sim 0.6$ kpc is the
estimated distance of the PSR J0737$-$3039 system \cite{lyne}.
The incident radio power is smaller by
\be\label{ratrad}
{P_{1.4~{\rm GHz}}\over P_{\rm wind}} =
{(\pi R_{\rm mag}^2) \nu F_\nu \over P_{\rm wind}} \simeq 
1.5\times 10^{-7}\,I_{A,45}^{-1}\,
\left({D_{\rm J0737}\over 0.6~{\rm kpc}}\right)^2.
\label{PrLA}
\ee

\subsection{Heating by External Radio Photons}

In spite of the small 
total radio power output, 
the absorption of the radio waves can have an important
influence on the kinematics of trapped electrons.
[See \cite{rafikov} for an independent and more detailed
analysis of this heating mechanism.]  For present purposes, 
we assume the existence of a flux of electrons (and possibly
positrons) moving trans-relativistically along the closed  
poloidal magnetic field line,
away from the star.  We now show that these particles will
be heated sufficiently to mirror and become trapped 
in the outer magnetosphere.  
The trapped particles are then further heated by the radio beam.


A non-relativistic electron absorbs (unpolarized) radio photons
with a cross section $\sigma_{\rm cyc}(\nu) = 
(\pi e^2/2m_ec)\,[1+ \cos ^2 \kappa]\delta(\nu-\nu_{B,e})$.
Here $\nu_{B,e} = eB/2\pi m_ec$ and
$\kappa$ is the angle between the direction of magnetic field
and the line of sight.   The energy absorbed by one particle
from the radio beam at a radius $r$ (outside the cooling radius
(\ref{rcool})) is
\be
{r\over v}{dE\over dt} \simeq {r\over v}\int \sigma_{\rm cyc}(\nu)
F_\nu\,d\nu,
\ee
when the particle motion is trans-relativistic.  This works out to
\ba
{r\over v}{dE\over dt} &=&  {\pi\over 4}
\,\left[1+\cos ^2 \kappa \right]\,
{er(\nu L_\nu)_{\nu_{B,e}} \over B(r) D_{AB}^2 v}\nn
&\sim&  400~\left[1+\cos ^2 \kappa\right]\,\left({v\over c}\right)^{-1}\,
\left({\nu_{B,e}\over 100~{\rm MHz}}\right)^{-4/3}\,
\left[{(\nu L^A_\nu)_{\nu_{B,e}}\over
10^{27}~{\rm ergs~s^{-1}}}\right]\;\;\;\;{\rm keV},\nn
\ea
when $1-\sqrt{1-v^2/c^2} = O(1)$.
This energy is absorbed far outside the cooling radius, and
so the particle mirrors soon after it begins to return to the
star.

Particles trapped in the magnetosphere will continue to be heated.
The rate of heating at any one position in the magnetosphere depends
on the column of intervening particles, and the pump spectrum.
Using  synchrotron absorption coefficient for
a thermal distribution of particles at low frequencies
we find that 
the time to heat the particles up to a temperature $T \gg m_ec^2/k_{\rm B}$ 
is short:
\be
\tau_{\rm heat} = {3k_{\rm B}T_e n_e\over \alpha_\nu (\nu F_\nu)_{\nu_p}} = 
74\,\nu_{p9}\,\left({k_{\rm B}T_e\over m_ec^2}\right)^{8/3}
\left[{(\nu F_\nu)_{\nu_p}\over 10^4~{\rm ergs~s^{-1}~cm^{-2}}}\right]^{-1}
\,\left({\nu_p\over\nu_{B,e}}\right)^{2/3}\;\;\;\;{\rm s}.
\ee
where $\nu_{\rm peak, 9}= \nu_{\rm peak} /10^9$ Hz.
Balancing optically thin synchrotron cooling at a rate
$16(k_{\rm B}T/m_ec^2)^2$
$(n_e\sigma_Tc)(B^2/8\pi)$ per unit volume, with synchrotron
heating at a rate $\alpha_\nu (\nu F_\nu)_{\nu_{\rm peak}}$ per unit volume,
one obtains an equilibrium Lorentz factor
\be\label{kteq}
{k_{\rm B}T_{\rm eq}\over m_ec^2}  = 
 9.6 \left( { B_{\rm NS} \over 5 \times 10^{11} {\rm G}} \right)^{-9/11}\,
\left[{(\nu F_\nu)_{\nu_{\rm peak}}\over 10^4~{\rm ergs~s^{-1}~cm^{-2}}}\right]^{3/11}\,
\left({\nu_{\rm peak}\over\nu_{B,e}}\right)^{-2/11}.
\ee
This is fairly close to the limiting temperature (\ref{ktmax})
obtained by balancing the particle pressure in the outer magnetosphere
with the magnetic pressure.  


\subsection{Electrostatic Heating and Thermalization of
an Optically Thick Plasma}
\label{bulkheat}

The wind energy of pulsar A that is incident on the magnetosphere of 
pulsar B will be converted, with some efficiency, to internally
generated radio waves.  Even if the radio output of the
magnetosphere $L_{\rm mag}$ is much weaker than that of 
pulsar A, its heating effect can dominate by the factor
$4(D_{AB}/R_{\rm mag})^2 
L_{\rm mag}/L_A \sim 2000 (R_{\rm mag}/4\times 10^9~{\rm cm})^{-2}
L_{\rm mag}/L_A$.  So we consider the possibility that the
magnetospheric plasma is, itself, a source of 
low frequency photons ($\nu < 100$ MHz) which are created and
absorbed in situ.   The unpulsed emission from the
PSR J0737$-$3039A/B system is several times brighter at 1400 MHz
than is the combined pulsed emission of the two neutron stars
\citep{burgay03}.  It is possible that some of this unpulsed emission
is generated in the magnetosphere of pulsar B.

Internal heating has another advantage over the absorption of
radio photons from an external pulsar:  it is more effective
when the seed particles move relativistically along the magnetic field.  
The freshly injected particles will only resonantly absorb photons 
of a very low frequency, $\nu \sim \gamma_\parallel^{-1}\nu_{B,e}$.
At this frequency, the external radiation sees a very large
optical depth (from the previously injected particles which have
already acquired large perpendicular energies).

When the density $n_e$ of electrons (and positrons) greatly exceeds
the corotation charge density, torsional waves generated
with a wavelength $k_\parallel r \sim 1$ will couple non-linearly
to  higher-frequency waves.
If the energy density in the light charges is still small compared
with that of the background magnetic field,
the inner scale of the resulting turbulent spectrum is determined
by balancing the fluctuating current density with the maximum
conduction current that can be supported by the plasma 
\citep{thomp98}.
  At high frequencies, the cascade
is expected on general grounds to be strongly anisotropic. 
The wavenumber $k_\perp$ of a wavepacket perpendicular to the
background magnetic field will be much greater than the parallel
component $k_\parallel$, and related to it by 
$k_\perp \delta B \sim k_\parallel B$ \citep{goldsh97}.
Balancing $k_\perp \delta B \sim 4\pi n_e e$, and relating
$n_e$ to the corotation charge density through
$n_e = \lambda_{\rm eq} (\Omega B/2\pi ec)$, gives
\be
k_\parallel r = 2\lambda_{\rm eq}\,\left({\Omega r\over c}\right).
\ee

Relativistic electrons moving anti-parallel to such 
an Alfv\'en wavepacket with Lorentz factor $\gamma_\parallel$
will emit photons of a low frequency (wiggler radiation),
\be\label{omwig}
\omega_A \sim \gamma_\parallel^2 k_\parallel c
\ee
\citep{fung04}.
Large parallel Lorentz factors are most easily achieved by
particles which are freshly injected with small perpendicular
energies into the outer magnetosphere, and then electrostatically
accelerated in regions of high turbulent intensity.  These
same charges will absorb the wiggler photons at their
rest-frame cyclotron resonance (lab frame frequency
$eB/2\pi m_e c \gamma_\parallel$) if
\be
\gamma_\parallel^3 \sim {1\over 2\lambda_{\rm eq}}\left({eB\over m_e c \Omega_B}\right)
 = {3\times 10^7\over\lambda_{\rm eq}}\,\left({B\over B_{\rm mag}}\right).
\ee
This condition is easily satisfied for the multiplicities
($\lambda_{\rm eq} \sim 10^5$) that we encountered in \S \ref{EM0}.
It is straightforward to check that the implied wiggler frequency
(\ref{omwig}) lies comfortably
above the plasma frequency at density $\lambda_{\rm eq}$.  It works out to
\be
{\omega_A\over 2\pi} = 3.0\,\left({\lambda_{\rm eq}\over 10^5}\right)^{1/3}\,
\,\left({B\over B_{\rm mag}}\right)^{2/3}\;\;\;\;{\rm MHz}.
\ee

The radio power needed to heat the freshly injected electrons is
a small fraction of the net power dissipated in the magnetosphere.
For example, if the particles carry a current
$J$ that supports a twist $\Delta\phi$ in the background magnetic
field, then their kinetic power (before electrostatic acceleration) is
\be
L_{\rm kin} \sim {\pi r^2 J\over e} m_ec^2
= 4\times 10^{23} \left({R_{\rm mag}\over 4\times 10^9~{\rm cm}}\right)\,
\left({r\over R_{\rm mag}}\right)^{-3}\;\;\;\;{\rm ergs~s^{-1}}.
\ee
The minimum radio power that is needed
to heat these particles is no larger than $L_{\rm kin}$.  
(It should, nonetheless
be emphasized that  some bunching of the radiating particles 
within the turbulent plasma is required to exceed this requirement.)  

Relativistic motion of the injected particles reduces the radio
power even further.  The particles
start in their lowest Landau states close to the star.
As they move out beyond the cooling radius (\ref{rcool}),
they will begin to decelerate as they absorb photons.  The
Lorentz factor parallel to the magnetic field is halved
when the energy of the absorbed photons in the particle
rest frame is $\Delta E \sim \gamma_\perp m_ec^2$.  
Here $\gamma_\perp = \sqrt{(p_\perp/m_ec)^2+1}$.  The energy
of the absorbed photons in the star's frame is smaller by
a factor $\sim \gamma_\parallel^{-1}$.  The net density
of the absorbed photons is therefore smaller than the
beam energy density,
\be
U_\gamma \sim {\gamma_\perp\over \gamma_\parallel} n_{\rm beam} m_ec^2
= {1\over \gamma_\parallel^2} U_{\rm beam} \propto \gamma_\parallel^{-1}.
\ee

The equilibrium electron temperature can be much higher in this
situation, than we found by assuming the external radio photons from
pulsar A to be the sole radiative pump.  
Balancing the rate of turbulent heating with the incoherent synchrotron output 
of the thermalized particles, and assuming that 
a fraction $\varepsilon_{\rm turb}$ of the wind energy density 
$B_{\rm mag}^2/8\pi$ is damped on a timescale $r/c$, one finds
\be\label{teqturb}
\lambda_{\rm eq}\,\left({k_{\rm B}T_e\over m_ec^2}\right)^2
= 2\times 10^{15}\,\varepsilon_{\rm turb}\,\left({r\over R_{\rm mag}}\right)^8
{\cal N}_B^{1/4}.
\ee
So, for example, if the energy density in torsional Alfv\'en waves
is a fraction $\varepsilon_A$ of the wind energy density at
a distance $r$ from pulsar B, then the three-wave damping timescale
is $\sim (\delta B/B)^{-2} (r/c) \sim \varepsilon_A^{-1} (B/B_{\rm mag})^2
(r/c)$.
One finds
\be
\varepsilon_{\rm turb} = \varepsilon_A^2 
\left({B\over B_{\rm mag}}\right)^{-2} = 
\varepsilon_A^2\,\left({r\over R_{\rm mag}}\right)^6.
\ee
The implied equilibrium temperature
is 
\be\label{tscale}
{k_{\rm B}T_e\over m_ec^2} 
= 230\,\varepsilon_A\,\left({\lambda_{\rm eq}\over 10^5}\right)^{-1/2}
{\cal N}_B^{15/8} \left( { r \over 1.5 \times 10^9 {\rm cm}} \right)^7
\ee
For $\varepsilon_A \sim 0.1$ this estimate is consistent with Eq. (\ref{ktmax}).



\subsection{Density Distribution  of the Heated Particles}
\label{bottle}

Suppose that at the edge of the magnetosphere, located at radius $R_{\rm mag}$, 
relativistic particles with a typical Lorentz factor $\gamma_0$ are 
isotropised.  Particles captured from the wind of pulsar A, and
particles injected into an optically thick plasma from below (\S
\ref{bulkheat}) are both expected to satisfy this criterion approximately.
Neglecting for the moment the effects of cyclo/synchrotron emission,
the pitch angle $\psi$ can be related to the value $\psi_{\rm mag}$
at a radius $r \sim R_{\rm mag}$ from the conservation of the first 
adiabatic invariant ${p_\perp^2 / B}$ and the constancy of
the Lorentz factor $\gamma_0$. [Here
$p_\perp = \sin\psi\,(\gamma_0^2 -1)^{1/2} m_ec$ is the 
is particle momentum perpendicular to the magnetic field.]
One finds
\be\label{mir}
\sin  \psi = \left({ B \over B_{\rm mag}}\right)^{1/2}\, \sin  \psi_{\rm mag}.
\ee
Reflection occurs when  $\psi = \pi/2$, where the field strength is
\be
{B_{\rm refl} \over B_{\rm mag}} = {1 \over \sin ^2 \psi_{\rm mag}}.
\ee
Near the axis of the dipolar field, the reflection radius is
$R_{\rm refl}/R_{\rm mag} \simeq  \sin ^{2/3} \psi_{\rm mag}$.
The particle spends only a short time $\sim R_{\rm refl}/c$ near
the reflection point, as may be seen from the solution
for the speed of the particle parallel to the magnetic field,
\be
v_\parallel = v_0 \cos\psi =
v_0 \left[1 - \left({B\over B_{\rm mag}}\right) \sin^2  \psi_{\rm mag} \right]^{1/2}.
\ee
Here $v_0/c = \sqrt{1-\gamma_0^{-2}}.$

The density of particles with initial pitch angle $\psi_{\rm mag}$ is
obtained from conservation of the particle flux,
\be
{dn \over d \psi_{\rm mag}} {v_\parallel\over  B} = 
{dn \over d \psi_{\rm mag}}\biggr|_{\rm mag}  {v_0\cos\psi_{\rm mag}\over B_{\rm mag}}.
\ee
Integrating over initial pitch angles up to a maximum value
$\psi_{\rm max} = \arcsin(B_{\rm mag}/B)^{1/2}$ gives the local density of particles,
\be
n =  2{dn\over d\psi_{\rm mag}}\biggr|_{\rm mag}\,\int_0^{\psi_{\rm max}}
{(B/B_{\rm mag})\cos \psi_{\rm mag}  \over 
\left[1 - (B/B_{\rm mag}) \sin^2\psi_{\rm mag} \right]^{1/2}}
\,{\sin \psi_{\rm mag}  d  \psi_{\rm mag}\over 2}
= n_{\rm mag}.
\ee
(The factor of 2 in front of this expression accounts for the
reflected flux of particles.)
Thus, the density of particles is approximately constant, in 
spite of the strong convergence of the magnetic field lines toward
smaller radius.  

The adiabatic invariant is no longer conserved if a particle reaches deep
enough into magnetosphere that it cools significantly.  This occurs
for electrons (and positrons) inside the radius (\ref{rcool}).  The
consequences for the equilibrium density of particles are discussed
in \S \ref{EM}.

\section{Geometrical Model of Eclipses}
\label{geom}

In this section we assume that closed field lines of within the
pulsar B magnetosphere are populated by relativistically hot plasma.  We 
calculate the synchrotron optical depth over a large number of lines of sight,
taking into account the three-dimensional structure of the magnetosphere.
We also assume, for simplicity, that the magnetic field is described
by a vacuum dipole, and that the absorbing plasma is truncating outside
some  radius.  This simple model can reproduce all
the salient features of the eclipse light curve, including an
essential part of the eclipse phenomenology, the weak 
frequency-dependence of the eclipse duration and dependence of phases of B.
The optical depth to synchrotron absorption
at all frequencies (especially at the  highest)
has a strong gradient near the eclipse boundary, and quickly becomes 
large ($\tau_\nu \geq 1$) over a small range of orbital phase.
This can be achieved if the absorbing particles are relativistic 
and can absorb in a wide frequency range.


\subsection{Synchrotron Absorption}

We assume that the closed field lines are populated by relativistic
electrons with a thermal distribution at
a temperature $k_{\rm B}T_e /m_ec^2 \simeq  10$ (see eq. [\ref{kteq}]).  
In this case, the peak of the synchrotron emission
is at a frequency  $\nu_{peak} \sim 4 (k_{\rm B}T_e/m_ec^2)^2\,\nu_{B,e}$. 
The cyclotron frequency is $\nu_{B,e}(B_{\rm mag}) \sim 2 \times 10^7$ Hz
at the edge of magnetosphere, and increases inward.
Therefore radio waves propagate  in the low frequency regime,
$\nu \ll \nu_{\rm peak}$, when the observing frequency is
in the gigahertz range.

The eigenmodes of the electromagnetic wave are linearly polarized
when ions are absent from the plasma.  The 
synchrotron absorption coefficients of the two
eigenmodes are then \citep{rybiki}
\be\label{K}
\alpha_\nu^{(1,2)} =
- {c^2 \over 8 \pi \nu^2} \int \epsilon^2 {d \over d  \epsilon} 
\left( { n(\epsilon) \over \epsilon^2} \right) P_\nu^{(1,2)} d \epsilon.
\ee
The polarization-averaged absorption coefficient is
\be\label{K2}
\alpha_\nu = {1\over 2}\left[\alpha_\nu^{(1)} + \alpha_\nu^{(2)}\right].
\ee
In these expressions,
the spectral power density emitted by a single particle is denoted by
\ba
P_\nu^{(1,2)} &=&
{ \sqrt{3} e^3 B \sin \kappa \over 4 \pi m_e c^2} 
\left[ \widetilde\nu \int_{\widetilde\nu}^\infty K_{5/3} (\eta) d\eta 
\;\pm\; \widetilde\nu\,K_{2/3}(\widetilde\nu)\right];\nn
\widetilde\nu &\equiv&  
\left({\epsilon\over m_ec^2}\right)^{-2}
{ 2\nu \over 3 (\sin \kappa) \nu_{B,e}};
\;\;\;\;\;\;\nu_{B,e} \equiv {eB\over 2\pi m_ec},
\ea
and we assume an isotropic distribution of pitch angles.  The
electromagnetic wave is absorbed on particles with pitch angle
equal to the angle $\kappa$ between ${\bf B}$ and the direction
$\hat k$ of wave propagation.
In polarization state (1), the electric vector is 
orthogonal to the ${\hat k}$-${\bf B}$ plane 
(the $+$ sign in eq. [\ref{K}]);  whereas
in polarization state (2), it sits in the ${\hat k}$-${\bf B}$ plane 
(the $-$ sign in eq. [\ref{K}]).
Also $\epsilon$ is the energy of the emitting particle,
$n(\epsilon)$ is particle distribution function, and
$K_{5/3}$ and $K_{2/3}$ are modified Bessel functions.

For a thermal distribution $n(\epsilon) \propto e^{-\epsilon/k_{\rm B}T}$ 
characterized by a temperature $T$ and total density $n_e$, 
the absorption coefficient below the peak frequency is
\ba\label{abcoeff}
 &&
\alpha_\nu =  {4\pi^2 e \,n_e\over 3^{7/3}  B \sin \kappa}
  \left[\left({m_ec^2\over k_{\rm B}T}\right)\,
{\nu_{B,e} \sin \kappa \over \nu }\right]^{5/3};
\nn
&& \alpha_\nu ^{(1)} = { 4 \over 3} \alpha_\nu; \;\;\;\;\;\;
\alpha_\nu ^{(2)} = {2 \over 3} \alpha_\nu .
\ea

\subsection{Eclipse Light Curve}

\begin{figure}[h]
\includegraphics[width=0.9\linewidth]{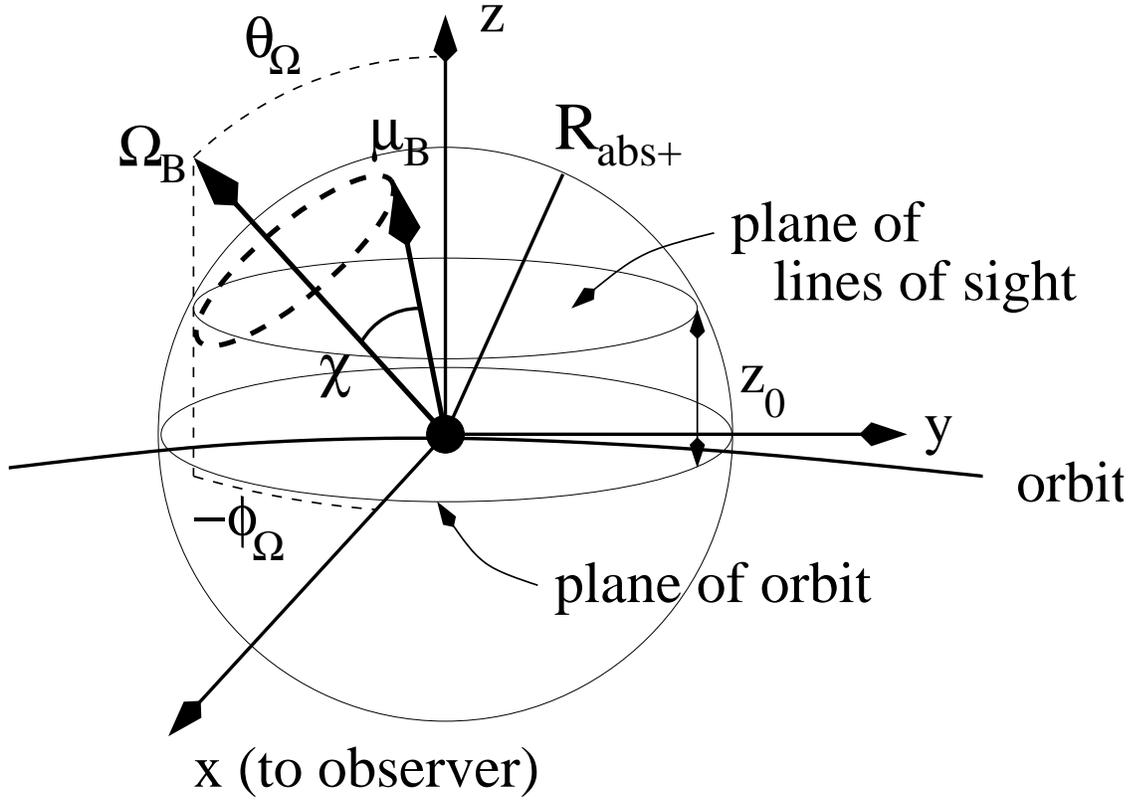}
\caption{Geometry of the model. The absorbing plasma is axisymmetrically
distributed about pulsar B, and extends as far as a distance
$R_{\rm abs+} \leq R_{\rm mag}$ from the neutron star.  
The spin axis of pulsar B is defined by angles 
$\theta_\Omega$ and $\phi_\Omega$.  The magnetic moment $\vec\mu_B$
makes angle $\chi$ with the  spin axis and rotates around ${\bf\Omega}_B$ 
every 2.77 s,
as the pulsar is carried by the orbital motion in the $y$-direction.
The line of sight passes at distance $z_0$ above the orbital plane.
}
\label{geom-new}
\end{figure}

We introduce a Cartesian system of coordinates $x,y,z$ centered 
on pulsar B (see Fig. \ref{geom-new}).  
The plane of the orbit is taken to coincide with the $x-y$ plane, 
from which the line of sight is offset by a vertical distance $z_0$ 
(eq. (\ref{zval})).
The observer is located at $x \rightarrow \infty$.
The spin axis of pulsar B is inclined at an angle 
$\theta_\Omega$ to the orbital normal, and at angle $\phi_\Omega$ 
with respect to the $x-z$ plane.  
The magnetic moment of pulsar B has a magnitude
$\mu_B = (B_{\rm NS} R_{\rm NS}^3)_B$, and is 
inclined at an angle $\chi$ with respect to ${\bf\Omega}_B$,
To calculate the intensity of the transmitted radiation, we need
to work out the  magnetic polar angle $\theta_\mu$ 
at each position ${\bf x} = [x,y(t),z_0]$ along the line of 
sight.\footnote{For the purposes of an initial calculation, we
neglect the possible presence of a toroidal magnetic field
(Appendix  \ref{Hpump}).}  Here
\be
y(t) = {\pi a\over 2}\left[1-\left({\phi\over 90~^\circ}\right)\right]
\ee
is expressed in terms of the semi-major axis $a $ cm
and the orbital phase $\phi$.  
The distance from a given point to pulsar B is
\be
r(x,t) = \sqrt{x^2 + y^2(t) + z_0^2}.
\ee
The magnetic polar angle of coordinate ${\bf r}$ is determined
from the unit vector $\hat {\bf \mu}(t)$ parallel to $\vec\mu_B(t)$,
\be
\cos\theta_\mu = {\hat{\bf \mu} \cdot {\bf r}\over r}.
\ee
The components of $\hat\mu$ are easy to write down in a
coordinate system aligned with ${\bf\Omega}_B$,
\be
\hat\mu_x^{\Omega} = \sin\chi \cos(\Omega_B t );
\;\;\;\;
\hat\mu_y^{\Omega} = \sin\chi \sin(\Omega_B t );
\;\;\;\;
\hat\mu_z^{\Omega} = \cos\chi.
\ee
Transforming to the observer's coordinate system gives
\be
\hat\mu_x = \left(\hat\mu_x^\Omega\cos\theta_\Omega +
                  \hat\mu_z^\Omega\sin\theta_\Omega\right)\cos\phi_\Omega
           - \hat\mu_y^\Omega\sin\phi_\Omega;
\ee
\be
\hat\mu_y = \hat\mu_y^\Omega\,\cos\phi_\Omega +
                \left(\hat\mu_x^\Omega\cos\theta_\Omega +
       \hat\mu_z^\Omega\sin\theta_\Omega\right)\sin\phi_\Omega;\;\;\;\;
\hat\mu_z = \hat\mu_z^\Omega\cos\theta_\Omega -  
                      \hat\mu_x^\Omega\sin\theta_\Omega.
\ee
The strength of the magnetic field at position ${\bf x}$ is then given by
\be
B={\sqrt{1+ 3 \cos^2 \theta_\mu} \over r^3} \mu_B.
\ee
The condition that a given point is located on closed field lines is that the maximum extension
of a field line is less than $ R_{\rm mag}$:
\be
R_{\rm max} = {r \over \sin^2  \theta_\mu } < R_{\rm mag}.
\ee
We also need to know
the angle between the line of sight and the local direction of the
magnetic field:
\be
\cos\kappa = {B_x \over B} = {3\cos\theta_\mu (x/r) - \hat\mu_x \over
                  (1 + 3\cos^2\theta_\mu)^{1/2}}.
\ee
For calculation of polarization properties,
the electric vectors of the two eigenmodes are parallel to the unit normals
\be
{\bf E}_1 = {B_y \hat z - B_z \hat y\over (B_y^2 + B_z^2)^{1/2}};
\;\;\;\;
{\bf E}_2 = {B_y \hat y + B_z \hat z\over (B_y^2 + B_z^2)^{1/2}}
\ee
which lie in the plane of the sky.  

We first calculate the total intensity of the transmitted
radiation.  Our examination of polarization effects is deferred to
\S \ref{Polariz}.  
The optical depth at time $t$, frequency $\nu$, and impact parameter
$z_0$ is obtained by integrating the polarization-averaged
absorption coefficient (\ref{abcoeff}) along the line of sight,
\be\label{tauint0}
\tau_\nu(t,z_0) = \int_{-\infty}^\infty \alpha_{\nu}\left[x, y(t), z_0
\right] dx.
\ee
We normalize the density $n_e$ of absorbing
particles to a characteristic Goldreich-Julian density at the
magnetospheric boundary ($B = B_{\rm mag}$):
\be\label{nedef}
n_e = \lambda_{\rm mag} n_{\rm GJ}(R_{\rm mag}) = 
\lambda_{\rm mag} {\Omega_B B_{\rm mag} \over 2\pi ec} = 
0.2 \,\lambda_{\rm mag}\;\;\;\;   {\rm cm^{-3}}.
\label{GJ}
\ee
(This is only a convenient normalization:  in fact, 
$\vec \mu_B$ and ${\bf\Omega}_B$ are nearly orthogonal in our best 
eclipse model.)  
We also normalize the line-of-sight coordinate and
field strength to $R_{\rm mag}$ and $B_{\rm mag}$ (eq. [\ref{Bn}]).
For our fiducial parameters, the optical depth is 
\be\label{tauint}
\tau_\nu(t,z_0) = 
     {4.5 \times 10^{-6} \over {\cal N}_B^{1/4}\nu_{\rm GHz}^{5/3}}
\int d \left( { x \over R_{\rm mag}}  \right) \,
\lambda_{\rm mag}\,\left({k_{\rm B}T_e\over 10\,m_ec^2} \right)^{-5/3}
 \left( { B  \sin\kappa \over B_{\rm mag}} \right)^{2/3}.
\ee
Here $\nu_{\rm GHz}$ is the wave frequency normalized to 1 GHz
and we have set $I_A = I_B = 10^{45}$ g cm$^2$.
In this expression, the parameter ${\cal N}_B$ encapsulates the
uncertainty in the magnetospheric radius through the normalization
of the spindown torque acting on pulsar B (eq. [\ref{Bn}]).
The code also tests whether the local cyclotron frequency satisfies 
condition $\nu_{B,e}/\nu >  \langle\gamma_e\rangle =
(3k_{\rm B}T_e/m_ec^2)$, and whether a  given
point is located outside of the cooling radius (\ref{rcool}).
Both of these effects are important only for extremely small impact 
parameters $z_0$.

As a simple prescription for density distribution we assume that
only those field lines are loaded with absorbing particles  for which
maximum extension (calculated at the magnetic equator) lies within a prescribed range,
\be\label{Rrange}
R_{\rm abs-} < R_{\rm max} < R_{\rm abs+}.
\ee
Motivated by \S \ref{bottle}, the plasma 
density $n_e$ and temperature $T_e$ are assumed to be
constant along each field line which lies in this range.  
We further assume that $n_e$ does not vary between field lines.

The onset and termination of the eclipse are then determined
by the physical boundary of the absorbing plasma.   We find
that $R_{\rm abs+}$ must be fixed at a value 
$\sim 1.5 \times 10^9$ cm in order to reproduce the width of
the observed eclipses.  The optical depth must quickly become 
large on lines of sight passing just inside the plasma boundary
(at a radius $\sim R_{\rm abs+}$).   The implied 
multiplicity is large, $\lambda_{\rm mag} \sim 10^5$, but still
consistent with our estimates in eqs. (\ref{multa}) and (\ref{multc}).

One expects a dipole to be only a rough approximation near the 
plasma boundary when $R_{\rm abs+} \sim R_{\rm mag}$.  However, the
dipole pressure rises rapidly inward, and so the dipole approximation becomes
increasingly accurate as $R_{\rm abs+}/R_{\rm mag}$ becomes smaller.
Modest deviations between the model and the data near the edges of 
the eclipse could be used to probe the distortion of magnetic field 
lines from a true dipole.

\subsection{Successful Eclipse Parameters}
\label{succ}

The calculation of the eclipse light curve requires a choice of
several parameters:  the angles $\theta_\Omega$, $\phi_\Omega$, 
$\chi$,  electron temperature $T_e$ and density multiplicity $\lambda_{\rm mag}$,
 inner and outer radii $R_{\rm abs\pm}$ and impact parameter $z_0$.
Our choice of the parameters was guided by a qualitative sense of
how they would influence the shape of the light curve, in the following
key respects:
\begin{itemize}
\item the eclipse is asymmetric, with the ingress shallower and longer than
the egress;
\item near ingress, transparent windows appear twice per rotation
of pulsar B;
\item near the center of the eclipse, the transparent windows
appear once per rotation;
\item near egress, the transparent windows are not well defined;
\item the duration of eclipse at a given rotational phase of pulsar B 
depends on the phase of B, being larger
when the magnetic moment is along the line of sight;
\item  the duration of eclipse at half maximum flux is $\sim 30 $ seconds, 
while the full duration is $\sim 40$ seconds.
\end {itemize}

\begin{figure}[h]
 \includegraphics[width=0.9\linewidth]{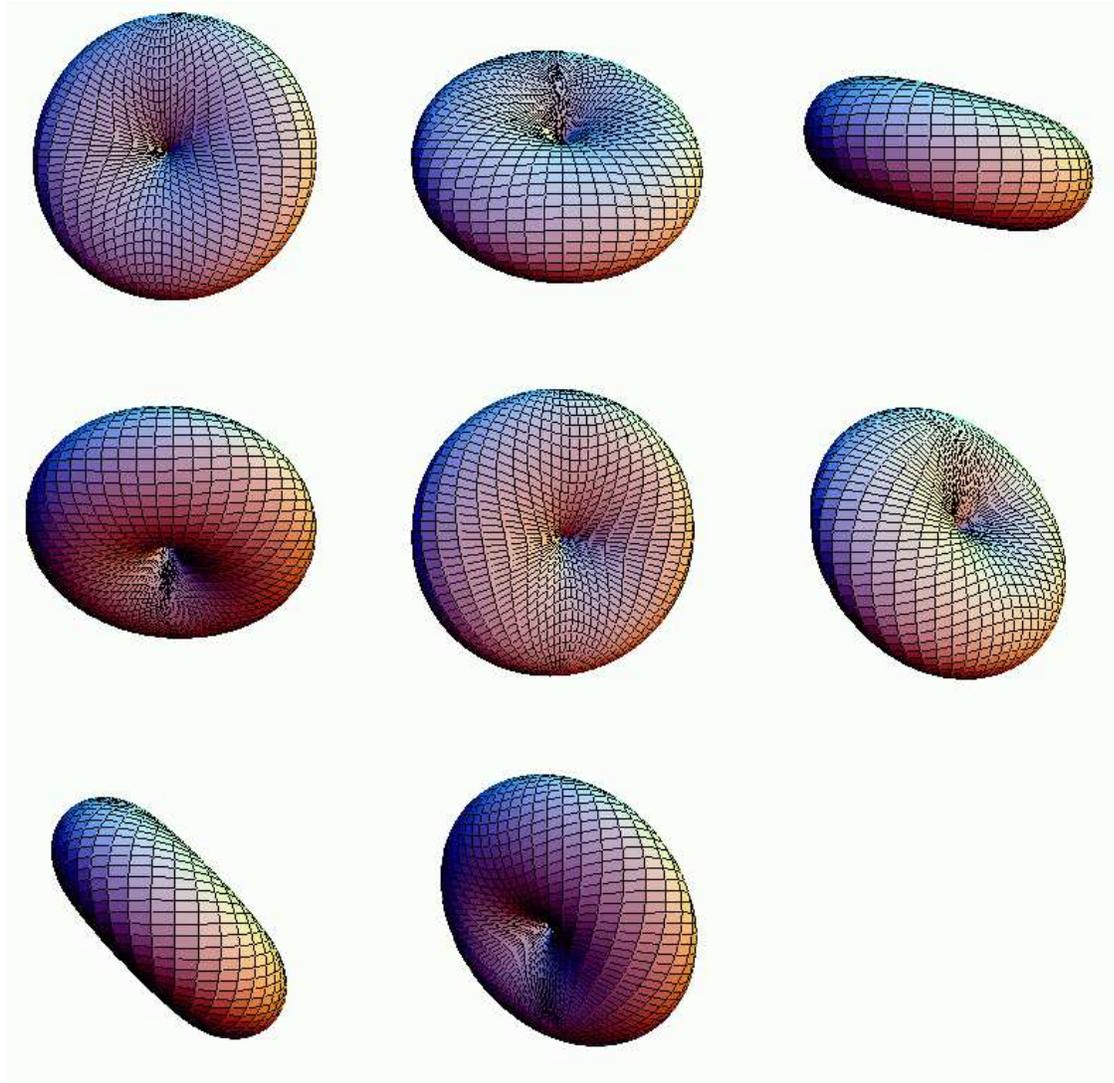}
\caption{View of the magnetosphere at different rotational phases
 separated by $\pi/4$.  For a full movie of the eclipse see 
 http://www.physics.mcgill.ca/$\sim$lyutikov/movie.gif}
  \label{movie}
  \end{figure}

\begin{figure}[h]
\includegraphics[width=0.95\linewidth]{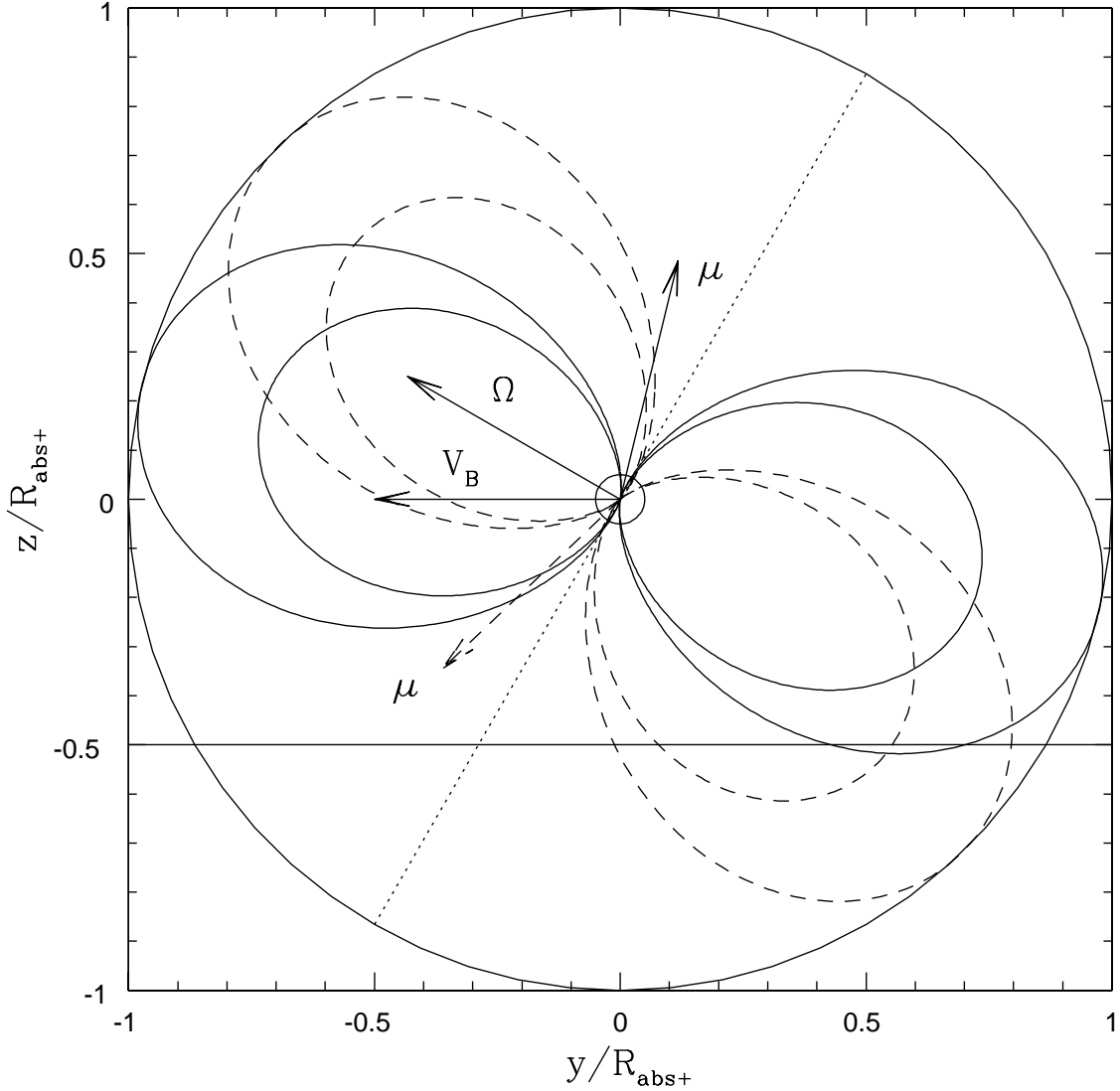}
\caption{Inferred geometry of the pulsar B magnetosphere projected 
onto the $z-y$ plane at the two  particular moments where
$\vec\mu_B$ lies in the $z-y$ plane.  The spin axis  ${\bf \Omega}_B$ is 
inclined by angle $\sim 60^\circ$ to the $z$ axis and lies in the $y-z$ plane.
The magnetic moment ${\bf \mu}_B$ makes an angle $\sim 75^\circ$ with 
${\bf \Omega}_B$. Apparent velocity of pulsar ${\bf V}_B$ is along 
$-y$ axis. The line of sight passes at $z_0/R_{\rm abs+} \sim -0.5$.
The absorbing plasma is located between  the two drawn magnetic surfaces
with $ R_{\rm abs-}= .75 R_{\rm  abs+} < R < R_{\rm  abs+}$. 
The inner circle marks the cooling radius $R_{\rm cool}=.05 R_{\rm  abs+}$.
The dotted line is the equatorial plane, orthogonal to  ${\bf \Omega}_B$.
The two directions of ${\bf \mu}_B$ plotted 
correspond to pulsar B phases separated by half a period.
Points corresponding to negative values of $y$ enter the line of sight first.
The transition to one transparent window per rotation occurs near the point
when the line of sight crosses the dashed field lines. 
A full eclipse occurs when the line of sight crosses both
solid and dashed field lines.}
\label{dipolepi6}
\end{figure}

The modulation of the radio flux during the eclipse is due to 
the fact that -- at some rotational phases of pulsar B -- the line 
of sight will only pass through open magnetic field lines where
is assumed to be negligible.
One of the main successes of the model is its ability
to reproduce both the single and double periodicities of these
transparent windows, at appropriate places in the eclipse.  
This requires that $\vec\mu_B$ be approximately -- but not
quite -- orthogonal to ${\bf\Omega}_B$.  To understand this result
geometrically, consider Figs. \ref{movie}-\ref{dipolepi6}.
The rotation of $\vec\mu_B$ out of the plane of the sky 
brings the observer's line of sight to within a small angle of
the magnetic pole (in the favored geometry).  
The cross-sectional area of the
absorbing plasma is maximized at this moment, and it projects
an ellipsoidal shape on the sky.
On the other hand, when $\vec \mu_B$ sits in the
plane of the sky, the absorbing plasma fills a region bounded
by a set of closed dipole field lines (Fig. \ref{dipolepi6}).
The shifting inclination of $\vec\mu_B$ through one half a rotation
of pulsar B allows the line of sight to intersect this dipolar
region once or twice.  The proportions of the eclipse
in which either periodicity is observed depend on the 
the impact parameter $z_0$ and the extent to which $\vec\mu_B$ is
nearly -- but not quite -- orthogonal to ${\bf\Omega}_B$.   A major
success of the model is that it can reproduce the relative
sequence and duration of these distinct periodicities.

\begin{figure}[h]
   \includegraphics[width=0.95\linewidth]{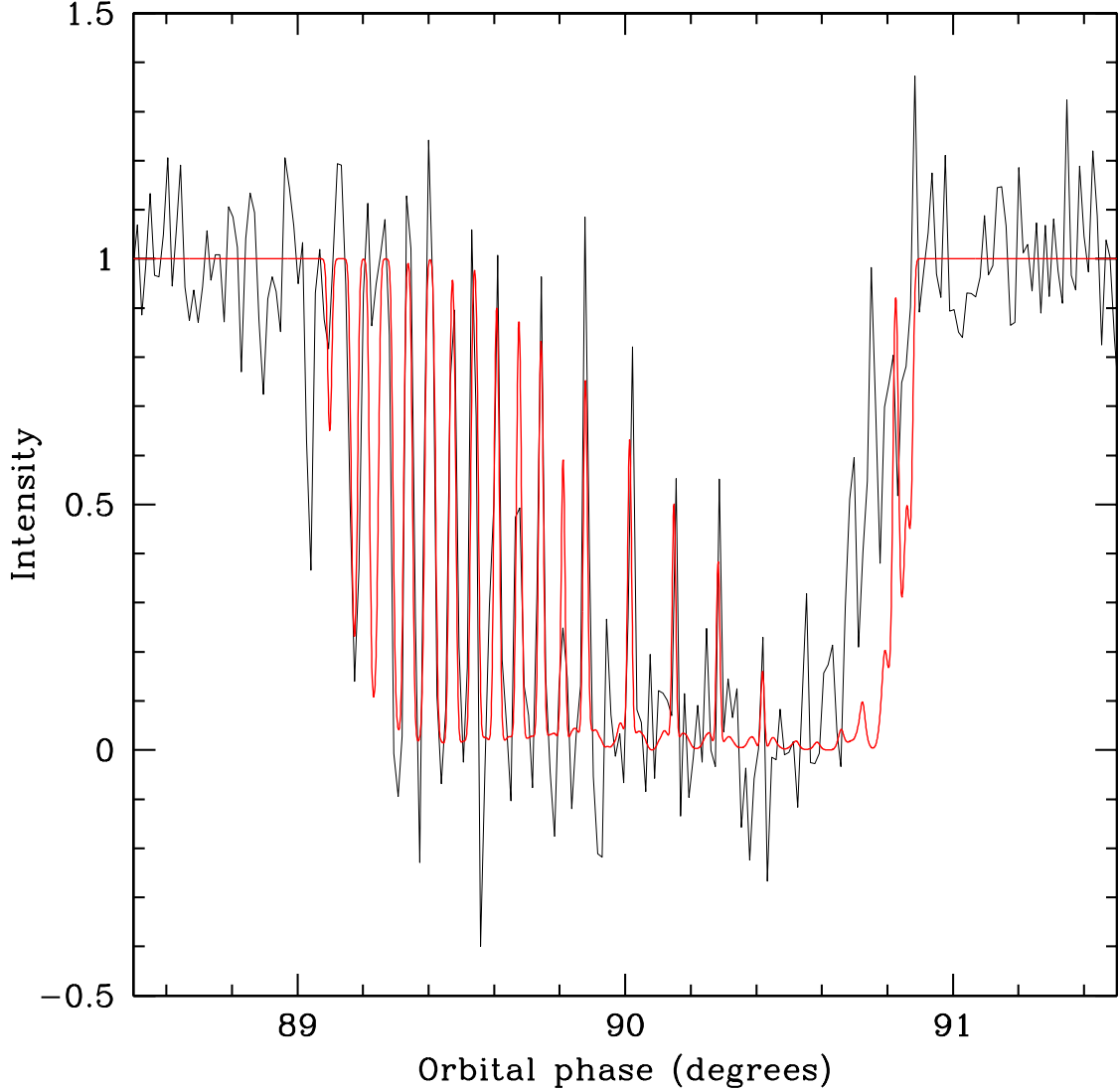}
   \caption{
    Comparison of a simulated eclipse profile (solid line) with 800 MHz data
    \citep{mcla04}.  In the model, the absorbing electrons are distributed
    on the dipole magnetic field lines with a constant density 
    $\lambda_{\rm mag} = 3\times 10^5\,(k_{\rm B}T_e/10\,m_ec^2)^{5/3}\,
    \nu_{\rm GHz}^{5/3}$ over a fixed range of radii $R_{\rm abs-} <
    R_{\rm max} < R_{\rm abs+}$.  Here $R_{\rm abs+} = 1.5\times 10^9$ cm
    and $R_{\rm abs-} = 0.75 R_{\rm abs+}$.   The same light curve will
    result if $\nu$, $T_e$ and $\lambda_{\rm mag}$ varied jointly so as to 
    keep $\lambda_{\rm mag}/(T_e\nu)^{5/3}$ fixed.  The spin axis of pulsar B
    is oriented at $\theta_\Omega/({1\over 2}\pi) = 0.66$ and $\phi_\Omega
    = -{1\over 2}\pi$.  The inclination between ${\bf\Omega}_B$ and the
    magnetic moment $\vec\mu_B$ is $\chi/({1\over 2}\pi) = 0.82$, and the
    impact parameter is $z_0 = -7.5\times 10^8$ cm.  The model light curve
    is smoothed over a timescale $\Delta t = 0.25$ s with a gaussian window 
    $e^{-\Delta\phi^2/2\sigma_\phi^2}$ [$\sigma_\phi=\Delta t/(2\pi)^{1/2}$].
    The model fits the data best in the middle of the eclipse.
 }
 \label{compare}
 \end{figure}

\begin{figure}[h]
  \includegraphics[width=0.95\linewidth]{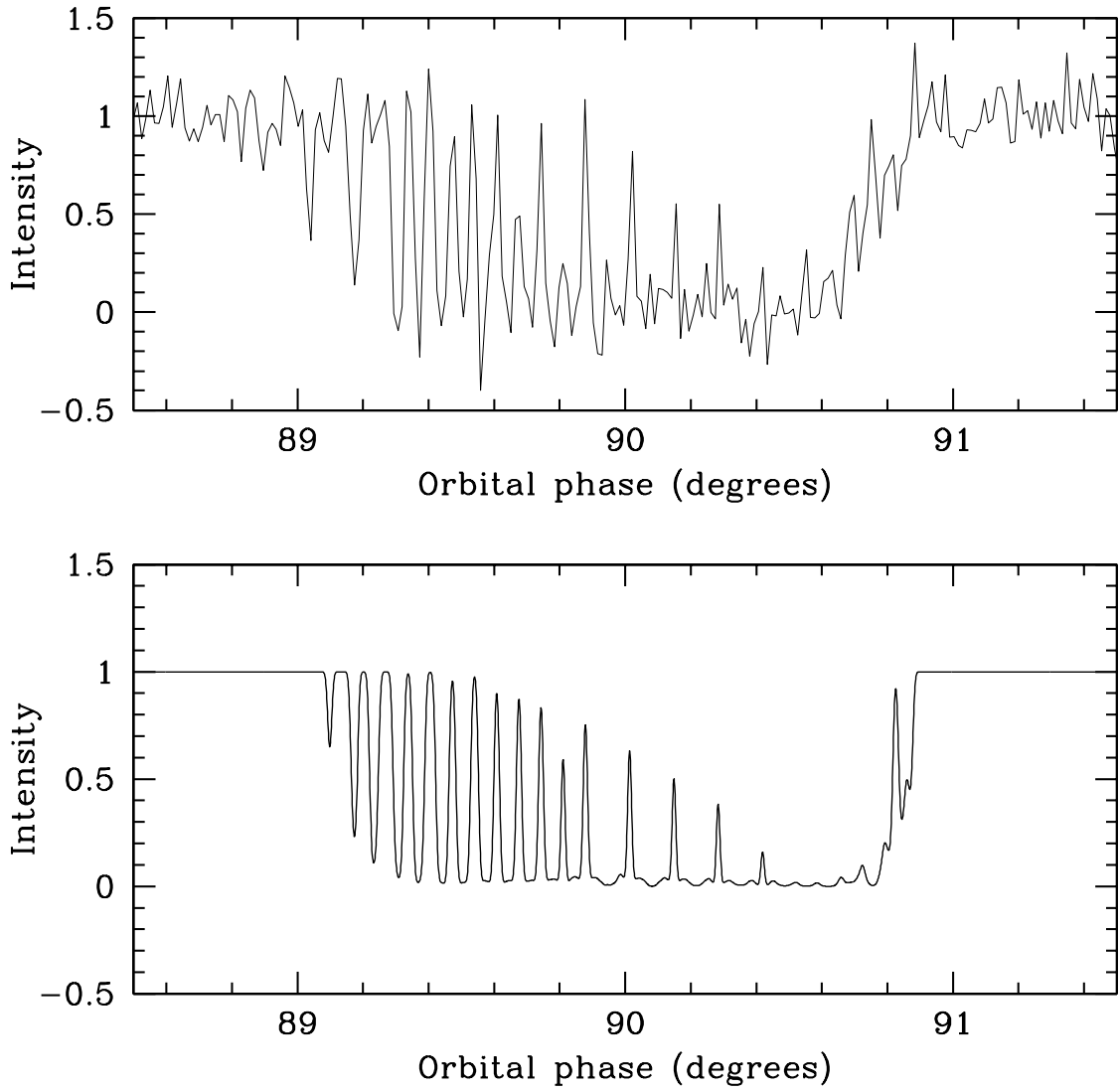}
  \caption{Same as Fig. \ref{compare}, but model and data side by side.}
  \label{compare1}
   \end{figure}

Our simulated light curve is displayed in Figs. \ref{compare} and  
\ref{compare1}.   Variations in the parameters of the model modify the 
light curve in the following ways.   A large asymmetry between ingress
and egress results from a combination of finite $\theta_\Omega$ 
(inclination between the spin axis of pulsar B and the orbital plane normal) 
and finite  $z_0$ (impact parameter).  The asymmetry is largely erased if
$\phi_\Omega$ is in the range $\pm 60^\circ$ (the rotation axis is out of
the plane of the sky).  There can be one or
two transparent windows per rotation of pulsar B,
depending on $z_0$ and $\chi$ (the angle between rotational 
angle and magnetic moment).  If $\chi$ differs considerably from
$\pi/2$, then both eclipse center and egress show strong 
modulation on a {\it single} rotational period.  
A very small value of $R_{\rm abs-}$ results in a complete eclipse in 
the center.  
Since $\nu \ga \nu_{B,e}$ in most or all of the eclipse, the inner 
boundary corresponding to $\nu_{B,e}/ \gamma_{\rm e} \nu > 1 $ is 
never reached.

Based on a sample of many eclipse light curves 
our best fit  parameters are:   
$z_0 \simeq -7.5\times 10^8$ cm,  
$\theta_\Omega\simeq 60^\circ $,
$\phi_\Omega \simeq -90 ^\circ$,  $\chi \simeq 75^\circ$.
The orbital inclination is therefore predicted to be close to
$90.55^\circ$. This is somewhat larger than that
inferred by \cite{coles} ($90.26^\circ \pm .13^\circ$), but consistent
with the estimate of  \cite{ransom04} ($88.7^\circ \pm .9^\circ$).  
Note that there is a symmetry in these parameters,
$z_0 \rightarrow -z_0$ and $\phi_\Omega 
\rightarrow -\phi_\Omega$,  which preserves the shape of the light curve.
In view of results of \cite{coles}, we have chosen
$z_0 < 0$. 

\begin{figure}[h]
  \includegraphics[width=0.95\linewidth]{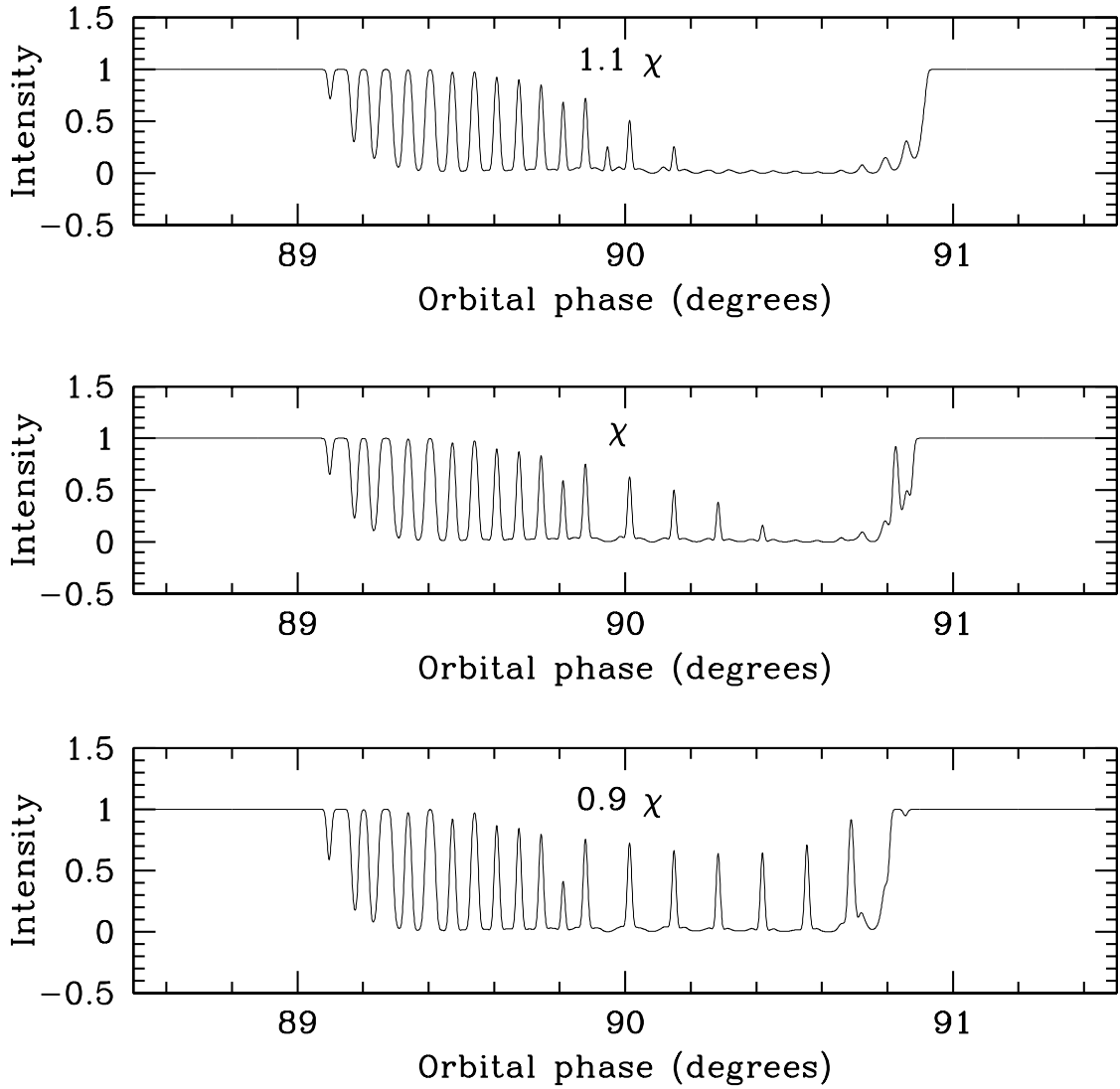}
   \caption{
    Change in the eclipse light curve due to a variation in the 
    impact parameter $z_0$ and the inclination $\chi$ between the
    magnetic and spin axes of pulsar B.
    Parameters are as in Fig. \ref{compare}.
 }
 \label{comparee}
 \end{figure}

The dependence of the eclipse profile on the 
parameters $z_0$, $\theta_\Omega$ and $\chi$ is illustrated in 
Fig. \ref{comparee}.  Overall the best constrained parameter is
$z_0 \sim -7.5\times 10^8$ cm, which must lie within a range
a range $\sim \pm 10\,\%$ of this value.  There is however, a degeneracy
between $\theta_\Omega$ and $\chi$ which produces nearly identical
light curves if both angles are varied up or down by the same factor.  
The results are not sensitive to $|\phi_\Omega + 90^\circ| \la 30^\circ$. 

The required {\it minimum}
multiplicity is 
$\lambda_{\rm mag} \simeq  3\times 10^5\,\nu_{\rm GHz}^{5/3}\,
(k_{\rm B}T_e/10\,m_ec^2)^{5/3}$ to reduce the transmitted
radio flux by $\ga 90\,\%$ at an orbital
phase of $\sim 90.5^\circ$.  This electron density 
is, in fact, close to the estimates (\ref{multa}) and (\ref{multc}).
The full duration of eclipse is about $40$ sec.
The size of the eclipsing region is $R_{\rm abs+} = 1.5\times 10^9$ cm,
so that the absorbing plasma must
be truncated well inside the expected radius of the magnetopause.
We comment on the significance of this result in the next section.

\begin{figure}[h]
  \includegraphics[width=0.95\linewidth]{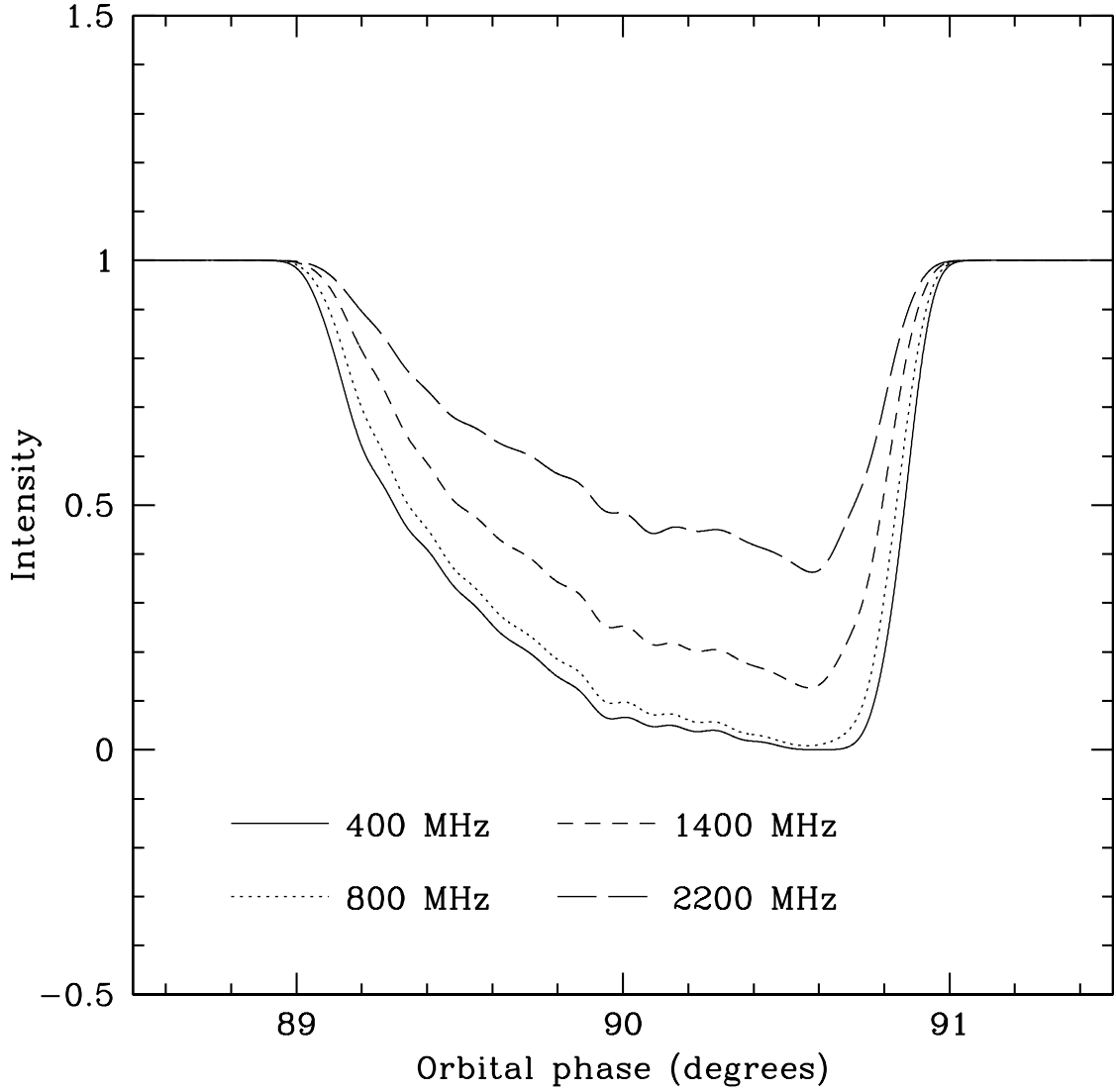}
  \caption{
      Frequency dependence of the eclipse profile, smoothed
      on a timescale $\Delta t = 2.77$ s.  Parameters are the same
      as in Fig. \ref{compare}.  At high frequencies eclipse duration should become 
      strongly frequency-dependent. This can be used to place tighter constraints on 
      plasma density.}
      \label{Ave}
    \end{figure}

\begin{figure}[h]
  \includegraphics[width=0.95\linewidth]{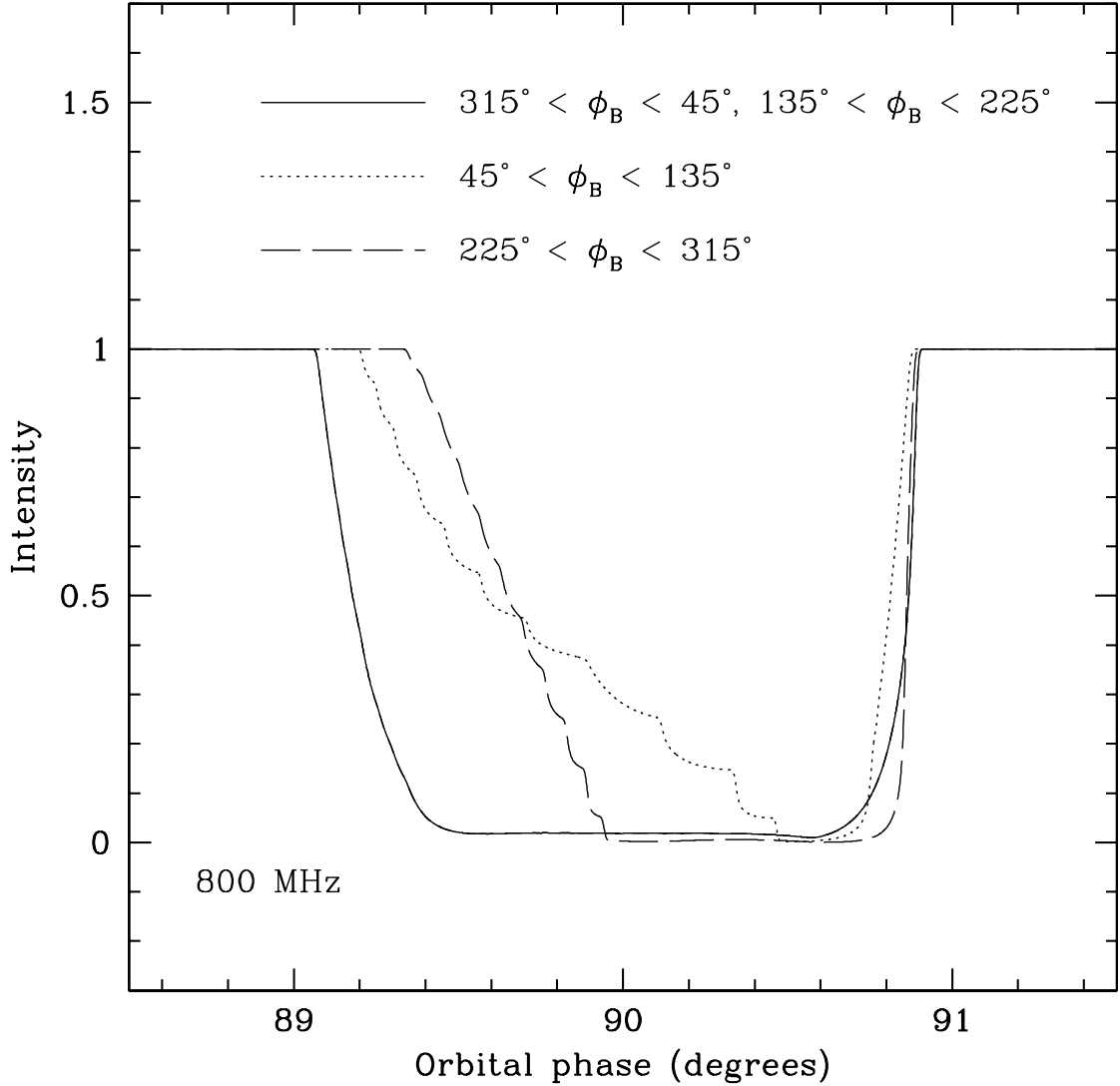}
  	\caption{Average eclipse profiles  for  different 
orientations of magnetic moment of B. 
Solid line: eclipse profile averaged over points 
$315^\circ<\phi_B < 45^\circ $ and $135^\circ < \phi_B < 225^\circ$, 
where 
$\phi_B$ is a rotational phase of B ($\phi_B \sim 0^\circ,\, 180^\circ$ 
corresponds
to magnetic moment pointing nearly towards an obsever, while
$\phi_B \sim 90^\circ,\, 270^\circ$ corresponds to magnetic moment
in the plane of the sky). 
Dotted line: eclipse profile averaged over  $45^\circ <\phi_B < 135^\circ$,
dashed line: eclipse profile averaged over 
 $ 225^\circ < \phi_B < 315^\circ$.
Eclipse is broader when the pulsar is ``face-on''.
     Parameters are the same as in Fig. \ref{compare}.}
 \label{eclipsalpha}
    \end{figure}

The model readily reproduces many fine details of the
eclipses.  It explains the modulation of that is observed at the first
and second harmonics of the spin frequency of pulsar B, and the deepening
of the eclipse after superior conjunction.
The average eclipse duration is almost independent of frequency
when the multiplicity is larger than $\lambda_{\rm mag} \sim 3\times 10^5$
(Fig. \ref{Ave}).  Eclipse is  broader  when
the magnetic moment of pulsar B is pointing closest to the observer,
just as is observed \citep{mcla04}.  
Figure  \ref{eclipsalpha} shows the 
eclipse profile averaged over different orientations of magnetic moment.
When $\vec\mu_B$  is pointing towards the observer the ``doughnut'' of the closed field lines
is seen nearly face-on, producing broad eclipses; when $\vec\mu_B$ is in the plane of the sky
the ``doughnut'' is seen edge-on,  producing narrower eclipses.

It should also be noted that
the details of eclipse ingress and egress
are not fully reproduced by the model:  in particular
the rise in the radio flux at egress is sharper than observed
(especially if $\lambda_{\rm mag}$ is high enough to give the eclipses
a weak frequency dependence).  
This could be due to deviations of the actual magnetic field from our
assumption of a pure dipole;  or be an artifact of our
assumption of a sharp plasma boundary.
We examine the effects of a smoother plasma profile and a variable
electron temperature profile in \S \ref{ions}.

\subsection{Eclipse Duration}

Our eclipse calculations show that the optical depth of the 
absorbing plasma undergoes a sharp drop at a distance  $R_{\rm abs+} \simeq
1.5\times 10^9$ cm from pulsar B.  
This is about $2.5$ times smaller than the expected radius
of the magnetopause, $R_{\rm mag} \simeq 4\times 10^{10}$ cm 
(assuming that the torque 
parameter ${\cal N}_B$ is unity in eq. (\ref{Bn})). The disagreement with the
maximum lateral extension of the closed field lines\footnote{Note that the
shock which decelerates the wind of pulsar A will have a yet larger
transverse dimension (by up a factor of 2 or so \citep{lyut,arons}.
This disfavors synchrotron
absorption in the shocked wind of pulsar A as the explanation for
the eclipse.}
is closer to a factor of 4 \citep[e.g.][]{lyut}.

The synchrotron optical depth (\ref{tauint}) reflects both the
geometrical distribution of absorbing particles, and their temperature
profile.  One possibility is that $R_{\rm mag}$ has been overestimated,
and that $R_{\rm abs+}$ reflects the actual boundary of the 
pulsar B magnetosphere.  Alternatively, the suppression in $\tau_\nu$
could be due to particle loss and overheating in the outer magnetosphere.
Let us consider these possibilities in turn (we clearly favor the latter). 

Recall that the torque parameter ${\cal N}_B$ is presently unknown.
The surface magnetic field of pulsar B could be smaller
than is implied by the the simplest estimate of the spindown
torque (${\cal N}_B \sim 1$; eq. (\ref{L2})). 
The torque formula
(\ref{L2}) assumes an increase in the opening of the dipole
field lines that is expected from simple geometry.  But other
effects are at work.  The pressure is distributed asymmetrically
about pulsar B, so that even if its spin were
in corotation with the orbit, it would be still be torqued by
the pulsar A wind \citep[the Magnus torque, e.g.][]{thomp94,arons}.
In addition, a strong surface resistivity at the interface between
the pulsar B magnetosphere and the shocked wind of pulsar A
would result in an enhanced spindown torque, due to the
resistive dragging of the poloidal field lines of pulsar B.  

There is a clear upper bound on the torque and spindown luminosity
of pulsar B:
\be
L_B \la   B_{\rm NS}^2 R_{\rm NS}^2  
c \left( { R_{\rm NS} \over  R_{\rm mag} } \right)^3 
{\Omega_B  R_{\rm NS}\over c}
\label{Lmax}
\ee
which comes from the limit of comparable toroidal
and  poloidal fields at the magnetospheric boundary.
 Eq. (\ref{Lmax}) implies that 
${\cal N}_B < c/( \Omega_B R_{\rm mag}) \sim 10$. 
If ${\cal N}_B$ saturated this bound, 
the  magnetospheric radius
$R_{\rm mag}$ would be  
\be
R_{\rm mag} =
\left({  I_B  \dot{\Omega}_B
\over I_A \Omega_A \dot{\Omega}_A}   \right)^{1/3}
  \left({ c D_{AB}^2 \over  2  } \right)^{1/3} =2.4 \times 10^{9} {\rm cm}
  \ee
The lateral radius of
the closed field lines is 50 percent larger, or $3.6\times 10^9$ cm,
which is still a factor of 2 too large. 
Thus,  one cannot explain small eclipse duration  as being {\rm entirely}
due to large torque/small magnetic field of pulsar B. 

The short eclipse duration may also reflect the loss of plasma from the 
outermost closed field lines, resulting from a change in the
topology of the field lines.  Experience with planetary magnetospheres
suggests that  the boundary between open and
closed field lines can vary by a factor $\sim 1.5$ in polar angle
(a factor $\sim 2$ in maximum dipole radius)
from the windward side to the leeward side \citep[e.g.][]{kabin04}.
Indeed, the relativistic PIC simulations of Arons et al. (in preparation) 
show that the plasma is very dynamic in the vicinity of the 
separatrix between open and closed field lines.

If the magnetic field in the pulsar A wind alternates in sign
over the pulsar A period (e.g. the wind is `striped')  then reconnection
can occur over most of the magnetopause surface.   On the other hand, 
if the direction of the wind magnetic field is constant in each
rotational hemisphere of pulsar A, then reconnection will occur
only on particular magnetospheric field lines that happen to run counter
to the wind field.  Trapped particles will, nonetheless, be 
effectively redistributed over the bulk of the magnetosphere by
azimuthal drift.  The drift
timescale for particles of Lorentz factor $\gamma_e$ 
is $\sim R_{\rm mag} ^2 \nu_{B,e}(B_{\rm mag}) 
/ \gamma_e c^2 \sim 2 \times 10^5$ s, which is shorter 
than the residency time $\sim (\Omega_B f_{\rm cool})^{-1} 
\sim 5 \times 10^5$ s (eq. [\ref{p}]).  As a result, the
loss of particles by reconnection in the magnetotail would
reduce the particle density on the windward side of the magnetosphere 
as well.

The loss of particles from the outer magnetosphere is likely to cause
overheating of the remaining particles, which reduces the 
synchrotron absorption coefficient (\ref{abcoeff}).   A similar
effect will occur if the trapped plasma has a large ion component,
since the density (\ref{neion}) of the electron-ion plasma 
is suppressed in the outer magnetosphere.  This variant of
the eclipse model is examined in the following section.

Two other explanations for a short eclipse duration can be discounted.
The angle between the line of sight and the
orbital plane could be larger than is claimed by 
\cite{coles} and \cite{ransom04}.  Increasing the 
inclination to $\sim 2^\circ$ would reduce the eclipse duration by a 
factor of two.  However, our modeling of the eclipse profile
places tight constraints
on the orbital inclination which disfavor this possibility.

Another unlikely explanation for the narrow eclipse width is
that the moment of inertia of pulsar B
-- but {\it not} of pulsar A (eq. [\ref{Bn}]) -- is much smaller than neutron
star models suggest.  More than an order of magnitude decrease in $I_B$ is
required, which is not possible for (the lower mass) 
pulsar B even if it were composed entirely of $u$-$d$-$s$ 
symmetric quark matter.


\section{Eclipses by an Ion-Supported Cloud}
\label{ions}

We now examine the case where the absorbing plasma is composed
largely of electrons and ions.  In this variant of the eclipse model, the
plasma is more centrally concentrated around pulsar B than was assumed
in \S \ref{geom}; and allowance
is made for a higher electron temperature in the outer magnetosphere.

The ions have two properties which
influence the plasma distribution.
First, they have a much stronger gravitational binding to the neutron star 
surface than do electrons and positrons; and, second,
they cool much more slowly in the outer magnetosphere.  
These properties of the ions have opposing effects on the 
equilibrium electron density in the outer magnetosphere.  On the
one hand, the large gravitational binding energy makes it 
energetically favorable
to force some of the slow-cooling electrons downward through the cooling
radius (with a parallel electric field) if their temperature is $k_{\rm B}T_e 
\la (A/Z)GM_{{\rm NS},B}m_p/R_{{\rm NS},B} \sim 200\,(A/Z)$ MeV.  Here
$Am_p$ is the ion mass, and $Ze$ its charge.

On the other hand, if the ions are themselves heated at the same
time as the electrons, they will prevent the lighter charges
cloud from collapsing through the cooling radius.  The principal
heating mechanisms which we examine in \S \ref{reabs}
involve the absorption of low-frequency transverse photons
(radio waves).  For example, heating of protons by such a mechanism will
be suppressed if the proton cyclotron frequency $\omega_{B,p} = 
eB/m_pc$ sits below the plasma frequency of the (relativistic) 
electrons, $\omega_{P,e} = (2\pi n_e e^2c^2/k_{\rm B}T_e)^{1/2}$.  
This gives a characteristic electron density
\be\label{neion}
n_e = \left({k_{\rm B}T_e\over m_pc^2}\right)\,{B^2\over 2\pi m_pc^2}.
\ee
The corresponding lower bound on the multiplicity of the 
neutralizing electrons (mass $m_e$) is
\be\label{multb}
\lambda_{\rm eq}^{e-i} = {180\over \cos\theta_B}\,
\left({k_{\rm B}T_e/m_ec^2\over 10}\right)\,
\left[ { (B_{\rm NS}R_{\rm NS}^3)_B 
            \over 5 \times 10^{29}~{\rm G~cm^3}} \right]\,
\left({r\over 4\times 10^9~{\rm cm}}\right)^{-3}.
\ee
This can be competitive with expression (\ref{multa})
somewhat inside the magnetopause, e.g. at $r \la 1\times 10^9$ cm.


From
the requirement that the electron pressure not exceed the magnetic
dipole pressure and using electron multiplicity (\ref{multb}),
the maximum temperature is 
\be\label{ktmaxb}
{k_{\rm B}T_{e,\rm max}\over m_ec^2} = {m_p\over 2m_e}\,
\left(1+{T_p\over T_e}\right)^{-1/2} = 9\times 10^2\,
\left(1+{T_p\over T_e}\right)^{-1/2},
\ee
which applies as long as $k_{\rm B}T_p \la m_pc^2$.


We therefore set the electron density to the threshold value
for significant ion heating.  Keeping in mind the results of
\S \ref{bottle}, $n_e$ is fixed at a constant 
value (\ref{neion}) 
\be
n_e(R_{\rm max}) 
= \left({k_{\rm B}T_e\over m_pc^2}\right)\,
{B_{\rm mag}^2\over 2\pi m_pc^2}\,
\left({R_{\rm max}\over R_{\rm mag}}\right)^{-6}
\ee
along each field line (labeled by the maximum radius $R_{\rm max}$).
The corresponding multiplicity parameter $\lambda_{\rm eq}^{e-i}$
is given by eq. (\ref{multb}) or, equivalently, by
\be
\lambda_{\rm mag} = \lambda_{\rm eq}^{e-i}(R_{\rm max})\,
\left({R_{\rm max}\over R_{\rm mag}}\right)^{-3}.
\ee
The factor of $(R_{\rm max}/R_{\rm mag})^{-3}$ in this expression accounts 
for the normalization of the multiplicity $\lambda_{\rm mag}$
at the magnetospheric boundary $R_{\rm mag}$.

One also expects a very strong dependence of $T_e$ on $r$
when the electrons absorb the energy of large-scale torsional
motions in the magnetosphere, and cool by incoherent synchrotron
emission (\SS \ref{bulkheat}-\ref{reabs}).
The equilibrium temperature of the absorbing electrons is 
expected to increase rapidly with radius (eq. [\ref{tscale}]).
The synchrotron optical depth given by eqs. (\ref{abcoeff}) and 
(\ref{tauint0}), with electron multiplicity (\ref{multb}), 
drops precipitously with distance from pulsar B,
\be\label{tauscale}
\tau_\nu(r) \;\propto\;  
r n_e B^{2/3} (\nu T_e)^{-5/3} \;\propto\; 
\varepsilon_A^{-4/9}\,\nu^{-5/3}\,r^{-97/9}.
\ee
(Here $\varepsilon_A$ is the energy density in torsional motions
at radius $r$ in the pulsar B magnetosphere, relative to the magnetic
pressure at the magnetopause boundary.)
One finds, generically, a sharp transition from large to small optical
depth, because of the much higher heating rates in the outer magnetosphere.
Taking $\varepsilon_A$ to be constant gives the frequency scaling
\be
\Delta\phi(\nu) \sim \nu^{-15/97}.
\ee

\begin{figure}[h]
  \includegraphics[width=0.95\linewidth]{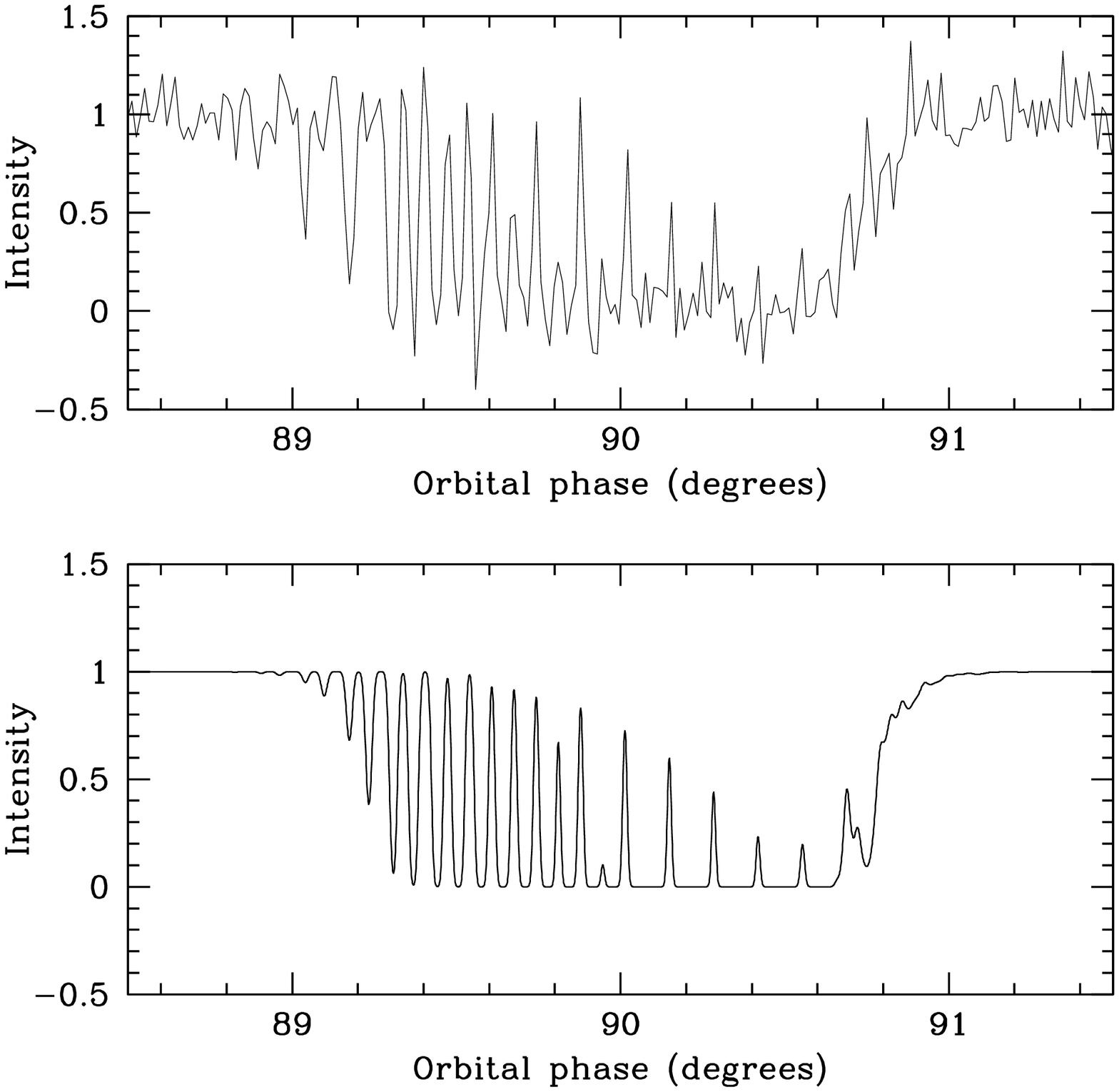}
   \caption{
    Comparison of a simulated eclipse profile with 800 MHz data
    \citep{mcla04} in the model with large ion content (\S \ref{ions}). 
    The density depends on radius according to Eq. (\ref{multb}),  
    and the temperature distribution is
     $k_{\rm B}T_e = 10\, m_ec^2\,\nu_{\rm GHz}^{-1}\, 
     (r/1.4\times 10^9~{\rm cm})^7 $.
     The magnetospheric radius is $R_{\rm mag}
    = 4\times 10^9$ cm (${\cal N}_B = 1$ in eq. [\ref{Bn}]).  The other
    parameters $z_0$, ${\bf\Omega}_B$, and $\chi$ are the same as
    in Fig. \ref{compare1}.
    The model light curve is smoothed over $\Delta t = 0.25$ s.
 }
 \label{compareb}
 \end{figure}

\begin{figure}[h]
  \includegraphics[width=0.95\linewidth]{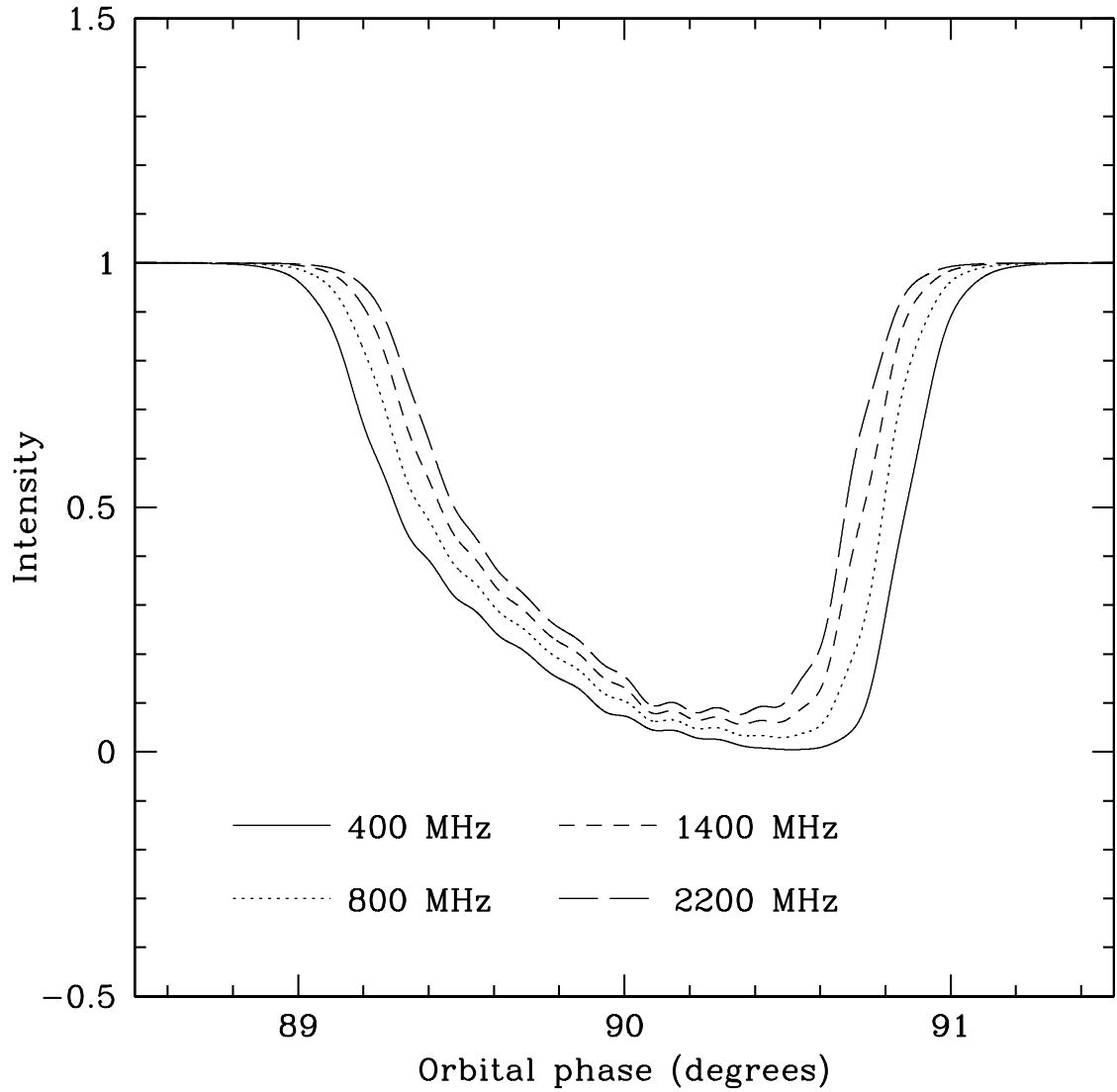}
  \caption{
      Frequency dependence of the eclipse profile in the model
      of Fig. \ref{compareb}, smoothed
      on a timescale $\Delta t = 2.77$ s.  Parameters are the same
      as in Fig. \ref{compareb}}
      \label{Aveb}
    \end{figure}

We therefore choose a temperature profile $T_e(r) \propto r^7$. 
After setting the magnetospheric radius to $R_{\rm mag} = 4\times 10^9$
cm (${\cal N}_B = 1$ in eq. [\ref{Bn}]), the only free parameter
in this model is the normalization of $T_e$, which is set to
$k_{\rm B}T_e = 10\,m_ec^2$ at $r = 1.4\times 10^9$ cm. 
The resulting eclipse light curves are displayed in Fig. \ref{compareb}.
One obtains an equally good fit to the central parts of the eclipse,
and a smoother fall and rise in the radio flux at ingress and egress.
The eclipse duration has a weak frequency dependence (Fig. \ref{Aveb}),
and the depth of the eclipse is less dependent on frequency than
in the constant-$n_e$ model (compare Fig. \ref{Ave}). 

A rapid increase in the temperature toward the edge of the
magnetosphere (combined with a drop in $n_e$ to maintain 
pressure balance) can therefore provide an explanation for
the duration of the observed eclipses.  Indeed, the dependence
of $T_e$ on $r$ could be even stronger than we assume if
 the amplitude $\varepsilon_A$ of the turbulence
 increases outward in the magnetosphere of pulsar B, and if
there is enhanced plasma loss from the outer magnetosphere.

It should also be noted that our model of electron heating involves
the absorption of low-frequency ($1-10$ MHz) radio waves
(\S \ref{bulkheat}).  This means that
$\tau_\nu$ will maintain the scaling (\ref{tauscale}), and
continue to drop off rapidly with radius, even
when the plasma is optically thin at GHz frequencies.

\section{Polarization}
\label{Polariz}

We now consider polarization effects associated with the 
propagation and absorption of the two electromagnetic modes in 
the eclipse region.  In a magnetized, relativistic plasma 
these modes are linearly polarized below the synchrotron peak frequency, 
$\nu \la \nu_{\rm peak} = \nu_{B,e} (k_{\rm B}T_e/m_ec^2)^2$,
even if the plasma is composed of electrons and ions
\citep[e.g.][]{sagiv04}.  The linear absorption coefficients
of the two modes are given by eq. (\ref{abcoeff}).  We present a
sample calculation of the polarization of the transmitted radio
pulses in our favored eclipse model, under the assumption that the
incident radiation is unpolarized.  In this case, the radiation
can become polarized as it propagates through an absorbing medium.

In an inhomogeneous plasma (where the
density and/or the magnetic field are functions of position),
it is more convenient to describe radiation transfer in terms of modes
polarized along fixed directions in space.  We will choose these
reference polarization states to lie along the normal to the orbital plane
($a$), and within the orbital plane ($b$).  The radiation field is then
characterized by  four Stokes parameters $I$, $Q$, $U$ and $V$.
The first two parameters can be re-expressed in terms of the intensities
in the two reference polarization states, $I = I^a + I^b$ and $Q =
I^a-I^b$.  The polarization fraction is $\Pi = \sqrt{U^2 +Q^2 + V^2}/I$.

Our calculation of the transmitted polarization neglects the 
effects of synchrotron re-emission and Faraday rotation.  The first
effect can be safely neglected if one is interested in the pulsed
component.  Faraday rotation is absent in a pair plasma, while for electron-ion
plasma at low frequencies it is suppressed by a factor 
$\sim \nu/\nu_{\rm peak}$ compared with the effects of synchrotron
absorption.  One consequence of this is that the effects of
mode tracking are also negligible.  This means that the instantaneous
polarization angle, $\chi = {1\over 2}\tan^{-1}(U/Q)$, 
is not able to adjust adiabatically to the changing direction
of the magnetic field \citep[e.g.][]{thomp94}.  

Marginally, effects of limiting polarization may lead to generation
of small circular  polarization, 
which depends on the 
 difference $n_2-n_1$ between the mode indices of refraction 
depends on the angle between $\hat k$ and ${\bf B}$.
In a relativistic  thermal plasma   $n_2-n_1$ 
depends on the angle between $\hat k$ and ${\bf B}$;
 approximately
\be\label{phshiftb}
(n_2-n_1){\omega\over c}  \;\simeq\; {\Gamma(1/3)\over
                                     2^{2/3}\pi}\,\alpha_\nu
\;=\; 0.54\,\alpha_\nu
\;\;\;\;\;\;\;\;\;\;
\left[\nu \ll \nu_{B,e}\left(k_{\rm B}T_e/m_ec^2\right)^2\right],
\ee
where $\alpha_\nu$ is the polarization-averaged 
synchrotron absorption coefficient \citep[e.g.][]{sagiv04}.
In this situation, the most important consequence of this phase shift 
is to generate finite circular polarization.  The circular polarization
parameter $V$ turns out to be small, but we include it in the calculation
for completeness.

Within the above approximations, the equations of polarization transfer 
read 
\ba 
{d I^a \over dx}
&=& - I^a \left[ \alpha_{\nu}^{(1)} \sin^4 \chi_B + 
\alpha_{\nu}^{(2)} \cos^4 \chi_B +  
{1\over 2}\alpha_{\nu} \sin^2(2\chi_B)\right]
+ {1\over 4} U \left[\alpha_{\nu}^{(1)}-\alpha_{\nu}^{(2)}\right] 
\sin(2\chi_B)
\nn
&&-{1\over 2}V\left[{\omega\over c}(n_2-n_1)\right]\sin(2\chi_B);
\nn 
{d I^b\over dx}
&=& - I^b \left[ \alpha_{\nu}^{(1)} \cos^4 \chi_B +  
\alpha_{\nu}^{(2)}\sin^4 \chi_B + 
  {1\over 2}\alpha_{\nu} \sin^2(2 \chi_B)  \right]
+ {1\over 4}  U 
    \left[\alpha_{\nu}^{(1)}-\alpha_{\nu}^{(2)}\right] \sin(2 \chi_B)
\nn
&&+{1\over 2}V\left[{\omega\over c}(n_2-n_1)\right]\sin(2\chi_B);
\nn 
 {d  U \over  dx}
 &=& - \alpha_{\nu}  U +  {1\over 2}(I^a+I^b) 
      \left[\alpha_{\nu}^{(1)}-\alpha_{\nu}^{(2)}\right]\,\sin(2\chi_B)
+ V\left[{\omega\over c}(n_2-n_1)\right]\cos(2\chi_B);
\nn
 {d  V \over  dx}
 &=& - \alpha_{\nu}  V 
+\left[(I^a-I^b)\sin(2\chi_B)-U\cos(2\chi_B)\right]{\omega\over c}(n_2-n_1)
\ea
\cite[e.g.][]{Pachol70}.
Here, $\chi_B$ is the angle between the reference direction $a$ 
and the projected magnetic field
(at a certain point $x$ along the line of sight).

\begin{figure}[h]
  \includegraphics[width=0.95\linewidth]{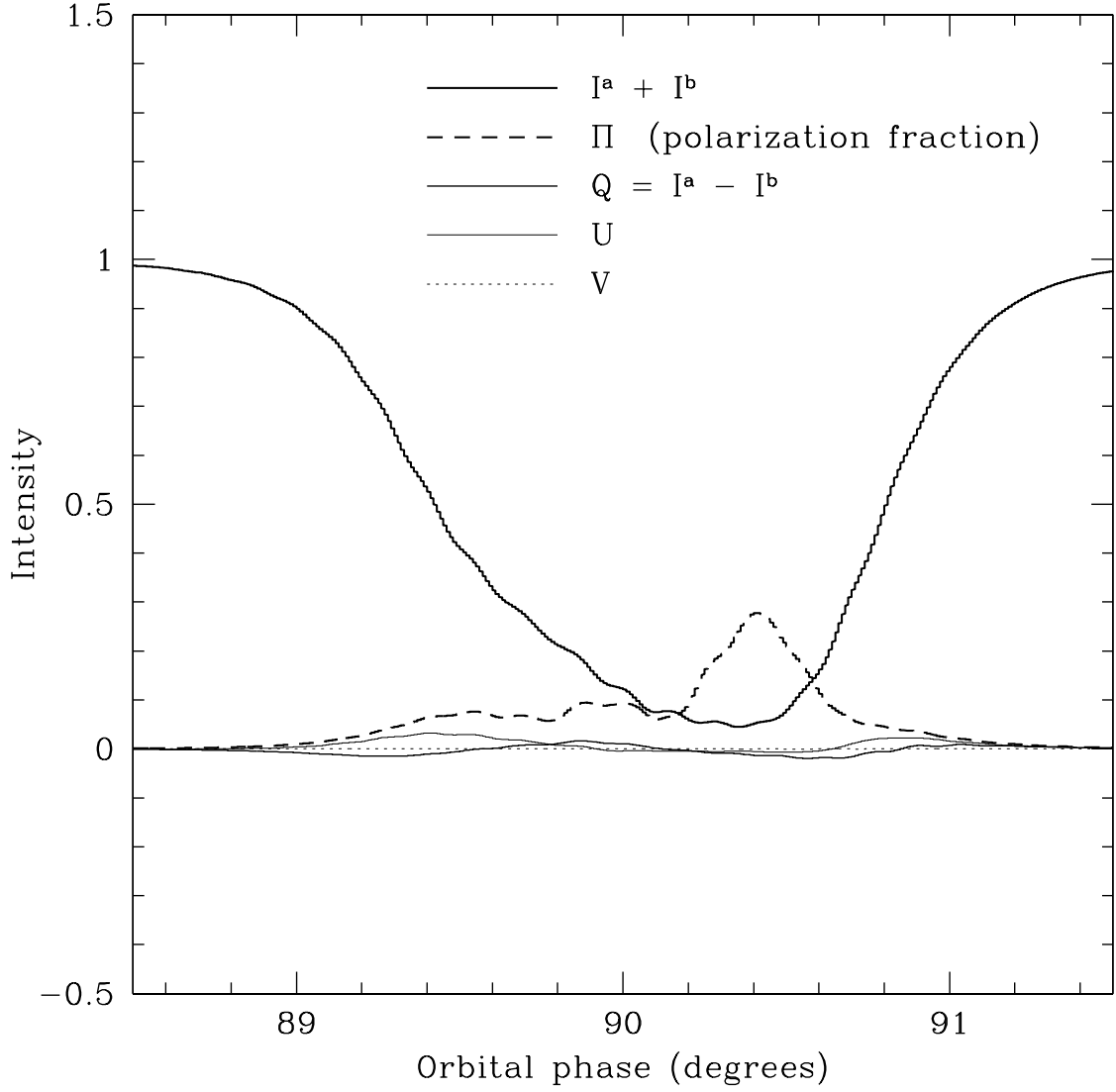}
\caption{Polarization generated during the eclipse of an unpolarized
background source, in the model of Fig. \ref{compare}.
 The reference directions
for the linear polarization are the  normal to the orbital plane
($a$); and the orbital plane ($b$).
Only negligible circular polarization is produced ($V\simeq 0$).
  }
 \label{TauPi0}
\end{figure}

The Stokes parameters are plotted in Fig. \ref{TauPi0} for our best fit
eclipse model.  The transmitted radiation is predicted to be strongly 
polarized in the deepest part of the eclipse, with a polarization fraction
reaching $25\%$. (This is, unfortuntely, the time when the signal to noise 
ratio is smallest.) The polarization angle evolves smoothly throughout eclipse.
One finds $ U \la Q$ when $\Pi$ is largest, 
which means that the transmitted polarization tends to
be aligned with one of the two reference polarization directions.
The main contribution to $U$ and $Q$ comes from
regions of the magnetosphere which have finite transparency.  

We find that the net value of $V$
is much smaller than that of $U$ and $Q$, by one or two
orders of magnitude, in the case where the background source is
unpolarized.  This is partly the result of a cancellation
between the ingoing and outgoing parts of the ray trajectory.



Polarization provides  an independent test of the eclipse mechanism 
and the geometry of the system.   The
radio pulses of pulsar A are, in fact, strongly polarized 
\citep[up to $50\%$:][]{ransom04}. 
In order to predict correctly the transmitted
polarization, one needs to know the direction of polarization
with respect to the orbital plane. 
This can be achieved using calibration of polarization
angle, \ie by referencing to a source with known position angle, 
and a determination of the orientation of the orbit on the sky (by
combining the scintillation measurements with the anticipated proper
motion measurements).

\section{Predictions and Comparison with Other Models }
\label{disc}

\subsection{Predictions of the model}

We now summarize the predictions of our model, and 
how further observations can be used to probe the geometry of the system
and the properties of the eclipsing material.

\begin{figure}[h]
  \includegraphics[width=0.95\linewidth]{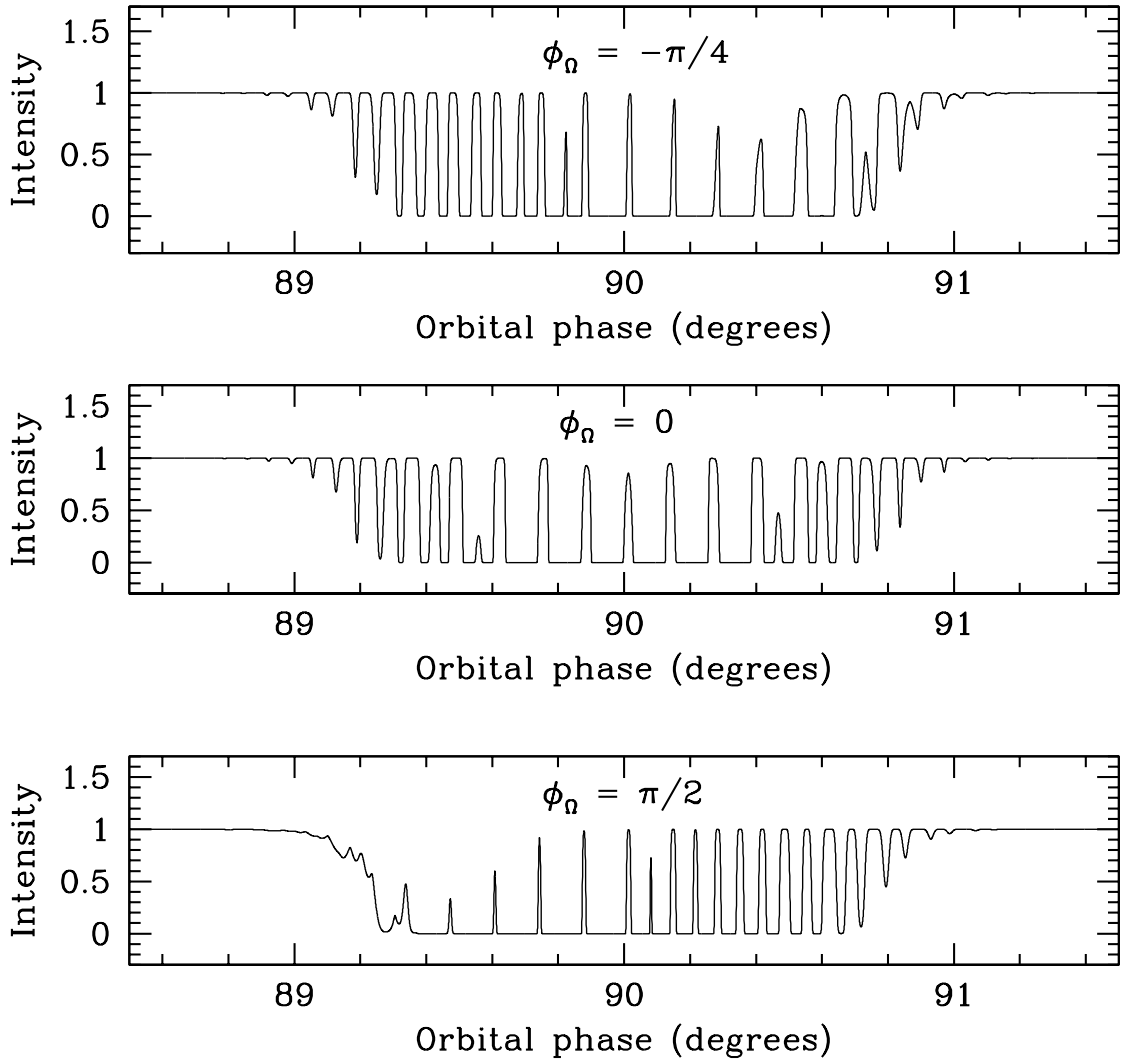}
   \caption{
    Prediction of the effects of geodetic precession on the 
    eclipse light curve.  The parameters are the same as in 
    Fig. \ref{compareb}, but with a temporal 
    resolution of $\Delta t = 0.01$ sec.
 }
 \label{compareg}
 \end{figure}

\begin{enumerate}
\item  The impact parameter $z_0$ between the line of sight and
the orbital plane is predicted to be $|z_0| = 7.5\pm 0.7\times 10^8$ cm.
The corresponding orbital inclination is $90.55^\circ\pm 0.05^\circ$
(or possibly $89.45^\circ\pm 0.05^\circ$).  This prediction is intermediate
between the measurements of \cite{coles} and \cite{ransom04}.

\item  The spin ${\bf\Omega}_B$ of pulsar B is expected to undergo
geodetic precession on a $\sim 75$ year timescale \citep{burgay03}.
We have provided predictions for how the eclipse light curve will
vary as a result.  In particular, the orbital phase at which the
radio flux reaches a minimum will shift back toward superior conjunction
(Fig. \ref{compareg}). Since $\phi_\Omega$ is not well constrained, 
a time for eclipse to become symmetric is between $\sim$ 12 years (if 
$|\phi_\Omega+90^\circ| \sim 30^\circ$ and ${\bf\Omega}_B$ is 
drifting  away from $\phi_\Omega=- 90^\circ$) 
and $\sim$ 25 years (if $\phi_\Omega$ is drifting toward $- 90^\circ$).


\item High time-resolution observations are a sensitive probe
of the distribution of plasma properties (density and temperature) on the
closed field lines. If plasma is depleted from  the outermost
field lines, at high temporal resolution the flux should return to unity.
On the other hand, if the absorbing plasma does not
have a sharp truncation in radius, then
the flux will not return to unity in all of the transparent windows.

\item  
The eclipses must regain 
a strong frequency dependence at sufficiently high frequencies.  The critical
frequency above which significant transmission occurs
can be used to place tight constraints on the plasma density.
The electron cyclotron frequency at a distance $r \sim z_0 \sim 7.5\times 10^8$
cm from pulsar B is estimated to be $\nu_{B,e} \sim 3$ GHz. 
One therefore expects the eclipses to develop a significant frequency
dependence at higher frequencies.

\item There are a number of
 definite predictions for the polarization of the transmitted
radiation as discussed in \S \ref{Polariz}.  The propagation
of the radio waves through the intervening magnetosphere is in
the quasi-transverse regime, where the polarization eigenmodes are
linear and Faraday rotation can be neglected.    The rotation
measure should therefore not change significantly during the eclipse,
and the variation in the dispersion measure is predicted to be 
small, of the order of $10^{13}$ cm$^{-2}$.  

\end{enumerate}
Finally, we note that the model predicts some radio emission from the
magnetosphere of pulsar B, but its precise amplitude and spectrum cannot
be easily determined from first principles.  A straightforward upper
bound to the radio output is obtained for incoherent synchrotron
radiation.  The luminosity of electrons (and positrons) radiating
at temperature $T_e$ can be no larger than the blackbody at this
temperature:
\be
{dL_{\rm bb}\over d\ln \nu} = 
{2\pi k_{\rm B}T_e \nu^3\over c^2}(4\pi R_{\rm mag}^2)
= 1\times 10^{22}\,\nu_{\rm GHz}^3\,
\left({k_{\rm B}T_e/m_ec^2\over 10}\right)\,
\left({R_{\rm mag}\over 4\times 10^9~{\rm cm}}\right)^2\;\;\;\;\;\;
{\rm ergs~s^{-1}}.
\ee
This is too weak to be important, and so the emission must be
of a coherent type such as we have described.

\subsection{Alternative Eclipse Models}

In this section we briefly contrast our favored eclipse model
with other mechanisms. 
At first sight, an
 important clue is that the electron cyclotron frequency,
measured along the line of sight, peaks at a value close to the 
observing frequency.   Thus, it is tempting to associate the
eclipses with  absorption of the radio waves on {\it non-relativistic}
particles.  Indeed, we have outlined a mechanism for supplying a constant
flux of non-relativistic electrons (and ions) on the closed magnetic
field lines (in Appendix \ref{Hpump}).  If the closed dipolar field lines
of pulsar B are twisted through an angle $\Delta\phi \sim 1$ radian,
then the optical depth through the cyclotron resonance (of either
type of charge) is $\tau_{\rm cyc} \sim \beta_\parallel^{-1}$. 
Here $\beta_\parallel$ is the drift speed of the charges, in units
of the speed of light.  Thus, a sub-relativistic drift of the
charges (e.g. with a speed $\beta_\parallel \sim \Omega r/c$)
would guarantee a significant optical depth at the cyclotron resonance.

There are two difficulties with this very simple model.  First,
the duration of the eclipse has a weak frequency dependence.
The eclipses extend out to a distance $R_\perp \simeq \pm 1.5\times
10^9$ cm from superior conjunction (in a direction parallel to the orbital
plane).  For example,
if the magnetic moment of Pulsar B were directed toward the observer,
then the maximum electron cyclotron frequency
along the line of sight would be
\be\label{nuceinb}
\nu_{B,e} = 400\;{\cal N}_B^{3/4}
\left({z_0^2\over R_\perp^2}+1\right)^{-3/2}\;\;\;\;{\rm MHz}
\ee
at eclipse ingress (or egress).   Here $z_0$ is the offset of
the line of sight from the orbital plane.  The inclination 
of the orbital plane with respect to the line of sight
is ${\pi\over 2}-i$, where $i$ is the inclination with respect
to the plane of the sky.  One then has
\be\label{zval}
z_0 = \left(i-{\pi\over 2}\right) D_{AB}
\ee
The corresponding expression is
\be\label{nucein}
\nu_{B,e} = 400\;{\cal N}_B^{3/4}
\left({z_0^2\over R_\perp^2}+1\right)^{-2}\,
\left({4z_0^2\over R_\perp^2}+1\right)^{1/2}\;\;\;\;{\rm MHz}
\ee
when the magnetic moment is oriented perpendicular
to the orbital plane.
Inverting these expressions, one obtains $R_\perp$ as a function
of $\nu_{B,e}$ for a fixed impact parameter.  In both cases,
it is a {\it stronger} function of frequency than
$R_\perp \sim \nu_{B,e}^{-1/3}$ when $b\neq 0$.
In addition, one observes that absorption must 
occur at a harmonic $> 1$ unless the torque parameter
${\cal N}_B > 1$ (see eq. \ref{Bn}).

Cyclotron absorption inside the magnetosphere
of pulsar B has some advantages:  the cross section
is larger than it is for synchrotron absorption by relativistic particles,
and the required particle density is thereby reduced. To have an appreciable 
optical depth at a radius $r$ (and corresponding frequency $\nu_{B,e}(r)$) 
the particle density should exceed the local Goldreich-Julian 
density by a factor $(\Omega_B r/ c)^{-1} \sim 10$.  Second,
the eclipse is observed to begin roughly where the line of sight
passes deep enough into the magnetosphere of pulsar B that
$\nu_{B,e} \sim \nu$.   Cyclotron absorption is, nonetheless, strongly
disfavored for at least two reasons:  first the expected frequency 
dependence of the eclipse duration is $\Delta\phi \propto \nu^{-1/3}$ (or a
stronger function of frequency), in contradiction with the
much weaker observed scaling  $\Delta\phi \propto \nu^{-0.1}$;
and, second, the absorbing particles will inevitably be heated
to relativistic temperatures.  In addition, a fluctuating component
of the current would cause a heavy fossil disk of cold particles,
which could be centrifugally suspended in the outer magnetosphere,
to be drained on a very short timescale \citep[cf.][]{tlk02}.

Our favored plasma parameters differ substantially from those
advocated by \cite{rafikov}, basically because we invoke a much stronger
heating mechanism, and a different source of relativistic particles.  
At high optical depth, particles are isotropized much more easily,
and the equilibrium particle density regulates to
a much larger value (due to the competition between injection
from below and precipitation through the cyclotron cooling
radius).  If the external radio photons are the primary heat source,
as advocated by \cite{rafikov}, then the optical depth at the cyclotron
fundamental is regulated to value of the order of unity.  One infers
that the eclipse duration should vary as $\Delta\phi \sim \nu^{-1/3}$
(or as a strong function of $\nu$ if the radio pump radiation is
brighter at $\sim 100$ MHz than it is at $\sim 1$ GHz).  

We have also considered effects of induced Compton scattering of 
pulsar A radio emission by plasma in  pulsar B magnetosphere and 
by pulsar B radio beam \citep[e.g.][]{thomp94}. Both
effects are negligible due to a small optical depth.

\section{Conclusion}
\label{concl}

We have developed a model of the radio eclipses of pulsar 
J0737$-$3039A, in which synchrotron
 absorption  on relativistic particles occurs inside the
intervening magnetosphere of the companion pulsar B.
We believe that the value of such modeling is threefold.
First, it provides a strong test of the longstanding assumption
that isolated neutron stars are surrounded by corotating, dipolar magnetic
fields.  Second, one is probing how a pulsar magnetosphere interacts
with an external wind, and in particular the mechanism by which
turbulence is damped in a relativistic and magnetically dominated
plasma.  Third, an understanding of how charges are supplied to
the magnetosphere may have broader implications for the electrodynamics
of radio pulsars.

Given the simplicity of the model, it is in excellent agreement with 
observations, especially in the middle of the eclipse where the
interaction with the external wind  has only a modest effect
on the shape of the poloidal field lines.  
The model can explain the asymmetry of the eclipse between ingress and
egress; the weak frequency dependence of its duration;
the modulation of the pulsar A emission at both the spin period of
pulsar B, and half its spin period, in different parts of the eclipse;
the detailed shape of the luminosity spikes in the middle of the eclipse;
and the dependence of the eclipse duration on the rotational
phase of pulsar B.   The model implies that pulsar B is nearly, but not 
exactly, an  orthogonal rotator.  It also requires the spin axis
of pulsar B to be tilted from the normal to the orbital plane.

We have demonstrated that -- at intermediate distances -- the poloidal magnetic
field of a neutron star is well approximated by a dipole.
This is a valuable confirmation of a fundamental assumption made in
models of pulsar electrodynamics. 
In  addition, our results are consistent with
models that place the source of the radio emission close to the magnetic axis.
Note that \cite{mcla04} define phases with respect to the arrival of 
radio pulses from pulsar B, whereas we define them with respect to the 
magnetic axis of B.   On the other hand, the dependence of eclipse duration
on pulsar B phase in our model is the same as inferred by \cite{mcla04}, 
so that the two definitions of phases are close to each other.

 In the present model we have neglected the fact that the radio pulse
profile of pulsar B has a single peak. 
This can be used to put addition constraints on the geometry
of the system, assuming that the radio emission mechanism 
is the same 
at both magnetic poles. In particular, in order for the radio beam from
one of the pole to miss the observer, the angle $\phi_\Omega $ should not
be equal to $\pm 90^\circ$. This comes from the fact that for $\phi_\Omega
=\pm 90^\circ$ the angle between the direction of the magnetic moment and the
line of sight is the same every half a period. Unfortunately, in the
absence of pulsar radio emission model with predictive power, this does
not provide any meaningful constraint at the moment.

Our eclipse model has implications for the
formation and evolution of the PSR J0737$-$3039A system.  The tilt
of the spin axis of pulsar B with respect to 
the normal to the orbital plane is consistent with a large natal kick 
(as has been inferred from the spatial velocity of the system: 
\cite{ransom04}). 
The kick would change the orientation of the orbital plane
and disrupt any pre-existing alignment between orbit and the
spin of the progenitor star.  Our measurement also does not support 
the suggestion of  \cite{demo04} that the spin axis should become aligned
with normal to orbital plane due to torque from pulsar A.

We have also described how an optically thick, relativistic, thermal plasma
may be maintained in the outer magnetosphere of pulsar B, where
synchrotron cooling is slow.  We have
argued that the interaction between the wind of pulsar A and the
magnetosphere is much more effective at heating the particles
than is the absorption of the radio pulses of pulsar A.  This
interaction also generates two sources of particles on the closed
magnetic field lines.  First, bunches of relativistic charges 
are created by a pulsar-type mechanism close to the neutron star
surface as the dipolar field lines are twisted back and forth close
to the magnetic poles.  Second, magnetic helicity builds up on the
dipolar field lines, as the result of the asymmetry between
the outgoing and return currents in each magnetic hemisphere.
The resulting steady current supplies a constant flow of electrons
and ions from the star to the outer magnetosphere.

The relativistic charges flowing in the magnetosphere of pulsar
B are generally a 
more potent source of radio photons than is the companion
pulsar.  The requirement for this to be true is that the
efficiency of conversion of turbulent energy to radio photons is 
larger than $\sim 10^{-7}$.  (For example, if the unpulsed radio
emission originates in the magnetosphere of pulsar B, then
this minimal efficiency is exceed by $\sim 10^4$.)
As a result, heating of the magnetospheric particles
occurs mainly through self-absorption of these internally generated
photons.   Heating of ions can have an important regulatory effect
on the equilibrium electron density:  if $n_e$ exceeds a critical
value, then radio photons that resonate with the ion gyromotion are
absorbed by the plasma.  A plausible source of low frequency pump
photons has also been identified.  This involves a two-step emission process:
a cascade of torsional Alfv\'en waves to high frequencies
where the waves become charge starved; followed by
wiggler emission of radio waves by the electrostatically accelerated
charges in the fluctuating magnetic field.

\begin{acknowledgements}
We would like to thank Maura McLaughlin for sharing original data and
comments on the  manuscript.  We also would like to thank Roman Rafikov,
Peter Goldreich, Jeremy Heyl, Victoria Kaspi,  Anatoly Spitkovsky, 
Ingrid Stairs, and Scott Ransom for stimulating discussions.
\end{acknowledgements}

{}

\appendix

\section{Helicity Pumping on the Closed Magnetic Field Lines}
\label{Hpump}

The excitation of large-scale torsional motions provides
a robust mechanism for drawing charges into the outer magnetosphere
of pulsar B. Damping of this turbulence will also heat the
suspended particles (\S \ref{bulkheat}), and if the heating
rate is too high then the plasma may not absorb radio waves
effectively.  This leads us to examine a subtler effect, 
involving a static component of the magnetospheric twist, which
will operate deeper in the magnetosphere of pulsar B where
the fluctuating current is small, 
$|\delta J| \ll |\rho_{\rm GJ}| c$. 

The exterior magnetic field of a neutron star can
store a considerable amount of magnetic helicity.  When it does,
the current flowing nearly parallel to the magnetic
field has a zero-frequency component.  In the magnetically
active Soft Gamma Repeaters and Anomalous X-ray Pulsars,
the unwinding of a strong, internal magnetic field is
a repeating source of magnetic helicity for the exterior
of the star \citep{tlk02}.   One does not expect such a mechanism
to be active in the much older (and weakly magnetic) PSR J0737$-$3039.

Nonetheless, we offer a simple argument why magnetic
helicity will be pumped from the outer to the inner magnetosphere
of pulsar B.  There are two simple components
to this argument.  First, if a fixed amount of helicity is injected
into a dipole magnetic field, then the energy in the toroidal magnetic
field is minimized if the helicity 
\be
{\cal H} \sim (\pi r^2)^2  B_P B_\phi \sim (\pi B_P r^2)^2 \Delta\phi
\ee
is concentrated close to the star \citep{tlk02}.
Here $B_P$ is the poloidal magnetic
field at radius $r$, and $B_\phi \sim B_P\Delta\phi$.  
The energy in the toroidal magnetic field is
\be
E_\phi \sim {1\over 12}B_\phi^2 r^3
\sim {{\cal H}^2\over 12\pi^2 r^5 B_P^2} \propto {\cal H}^2 r.
\ee
If a magnetosphere carrying helicity ${\cal H}$ is not in
the minimum energy state, the winding of the magnetic field
can be rapidly transferred by reconnection between different flux surfaces
(e.g. Taylor 1974).

The second observation is that the current flowing out along
an open bundle of magnetic field lines will not be compensated
precisely by the return current in the same magnetic hemisphere.
There is, as a result, a residual current connecting the two magnetic
poles, which flows along the outermost closed magnetic field lines.
This closed field current imparts a small amount of helicity
to the magnetosphere.  By the above argument, this helicity will
be pumped, by reconnection, on a short timescale $\sim c/\Omega$
to the inner magnetosphere.  The net rate of transfer of helicity
is then
\be\label{hplus}
{d{\cal H}\over dt}\biggr|_+ \sim \varepsilon_{\cal H}
\Phi_{\rm open}^2 \Omega.
\ee
Here $\Phi_{\rm open}$ is the magnetic flux $\Phi(r) = \pi B_P(r)r^2$
evaluated at the boundary radius of the magnetosphere.  

In the case of an isolated neutron star, an imbalance between the 
outgoing and return current can result from an offset
$\Delta R_{\rm dipole}$ of the magnetic dipole from the center of
the star.  One then deduces
\be
\varepsilon_{\cal H} \sim {\Omega \Delta R_{\rm dipole}\over c}
 < {\Omega R_{\rm NS}\over c}.
\ee
Making use of $\Phi_{\rm open} = \pi B_{\rm pole} R_{\rm NS}^2
(\Omega R_{\rm NS}/c)$, one can relate the helicity increase
(\ref{hplus}) to the magnetic dipole luminosity
$L_{\rm MDR} = {1\over 6} (\Omega R_{\rm NS}/c)^4
B_{\rm pole}^2 R_{\rm NS}^2 c$,
\be
{d{\cal H}\over dt}\biggr|_+ \sim 6\pi^2 L_{\rm MDR} \Delta R_{\rm dipole}.
\ee
where $B_{\rm pole}$ is magnetic field at the magnetic pole of the star.

The effect is much stronger in the
case of PSR J0737$-$3039B.  The strong asymmetry in the external
stress of the pulsar A wind will enforce a large asymmetry between
the outgoing and return currents in each magnetic hemisphere.
We therefore estimate
\be\label{varh}
\varepsilon_{\cal H} \sim \Delta\phi(R_{\rm mag}) \sim 
{\Omega_B R_{\rm mag}\over c} \sim 0.1-0.2.
\ee

The equilibrium twist angle that is established at smaller
radii depends on how rapidly the helicity is damped in the
inner magnetosphere.  We will not examine this problem here, and
simply note that the inward flux of helicity allows
one to establish a minimal static twist angle $\Delta\phi$ in
the outer parts of the magnetosphere.  Given that the twist is transferred
between flux surfaces with a speed $\sim c$,
one has
\be
{d{\cal H}\over dt}\biggr|_+ \sim \Phi^2(r)\,\Delta\phi(r)\,
{c\over r}.
\ee
Combining this with eqs. (\ref{hplus}) and (\ref{varh}) gives
\be
\Delta\phi(r) \sim 
\left({\Omega_B R_{\rm mag}\over c}\right)\,
\left({r\over R_{\rm mag}}\right)^3.
\ee
The corresponding current is
\be
{J\over |\rho_{\rm GJ}|c} \sim 3\,\left({\cos\theta_B\over 0.3}\right)^{-1}\,
\left({r\over R_{\rm mag}}\right)^2.  
\ee
It will be noted that the presence of a zero-frequency component
of the current will
influence the non-linear couplings between torsional
waves in the magnetosphere \citep{goldsh97,thomp98}.

\vfil\eject

 \begin{figure}[h]
  \includegraphics[width=0.95\linewidth]{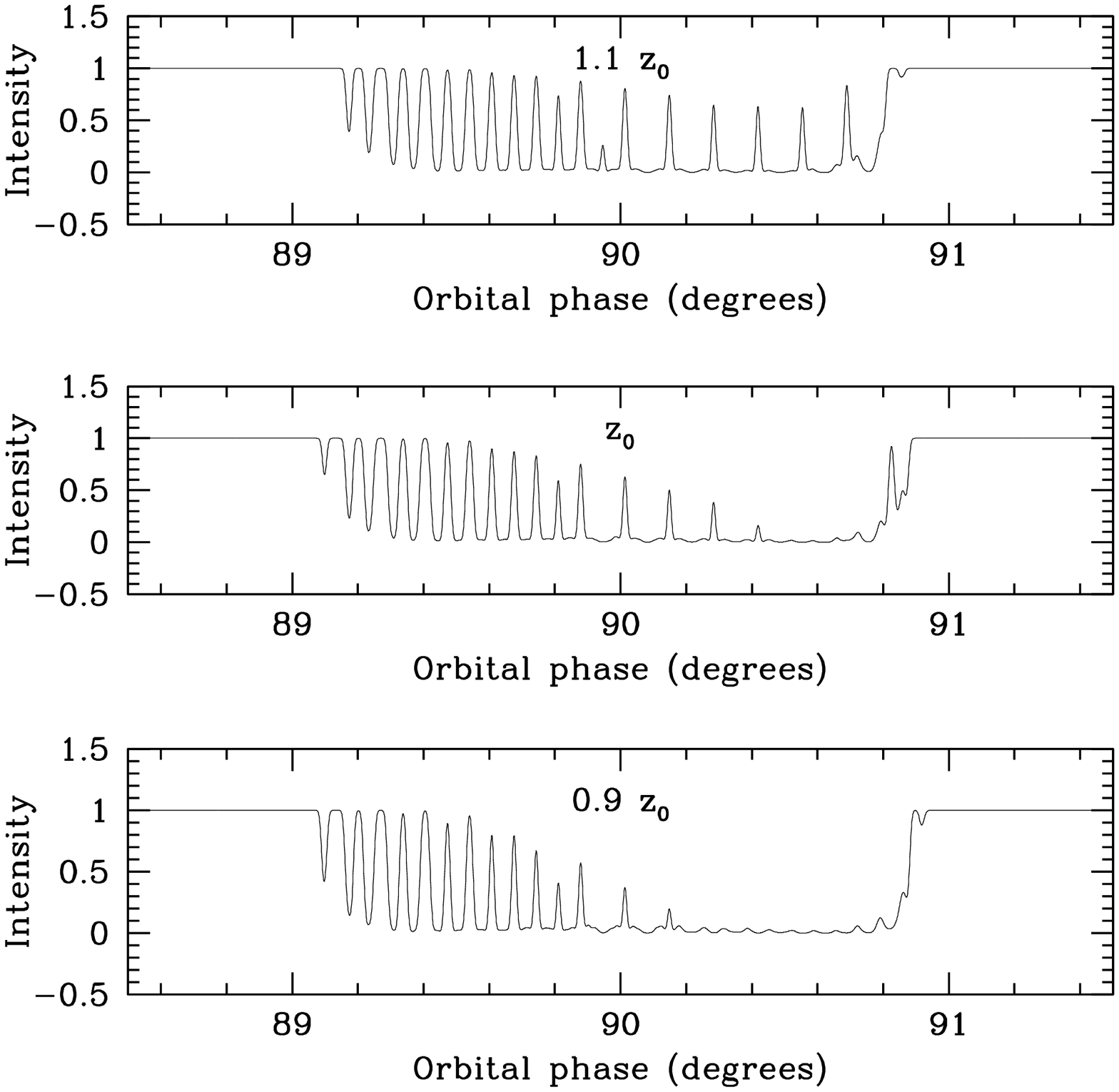}
 \end{figure}

\begin{figure}[h]
\includegraphics[width=0.45\linewidth]{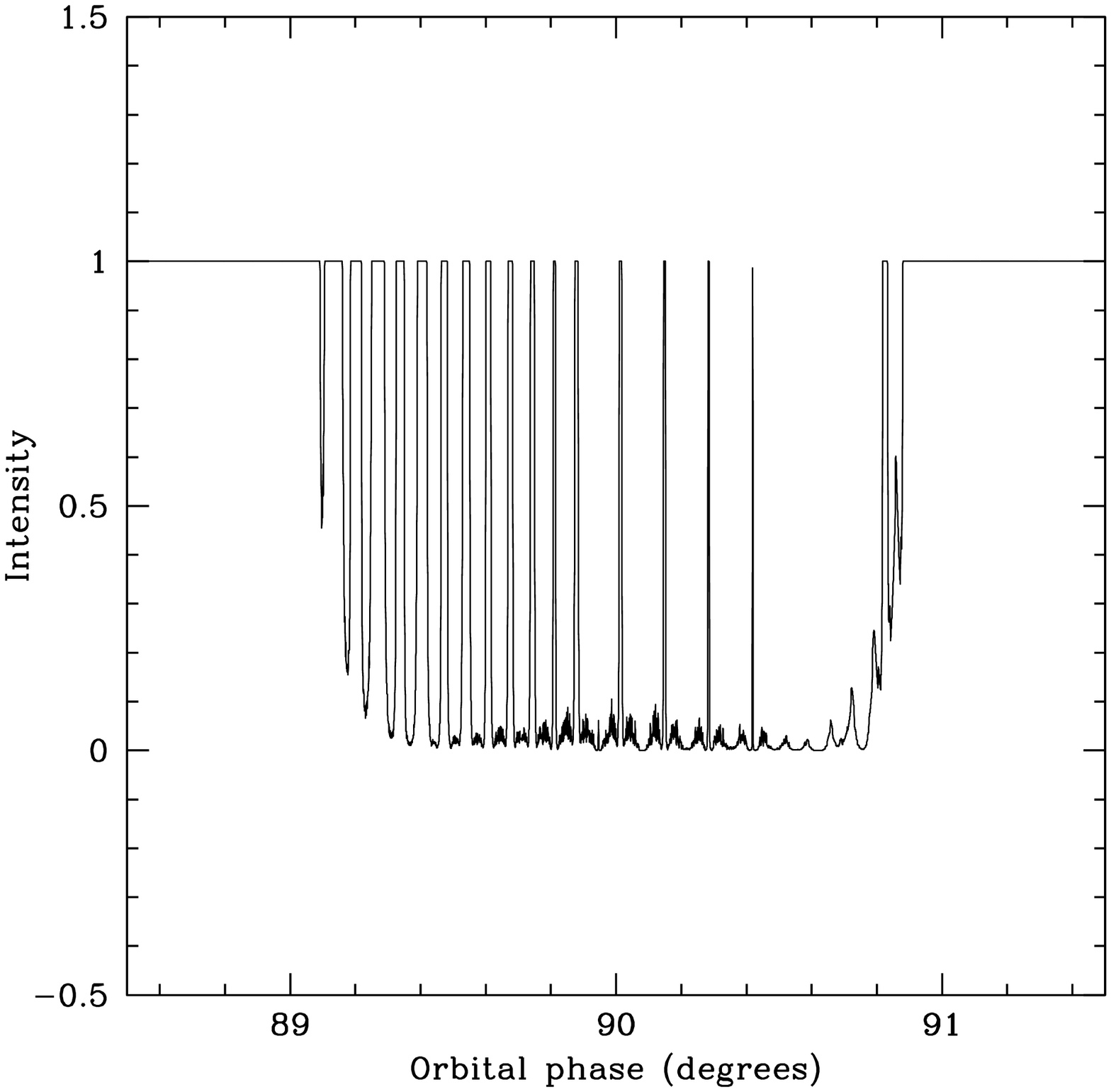}
\includegraphics[width=0.45\linewidth]{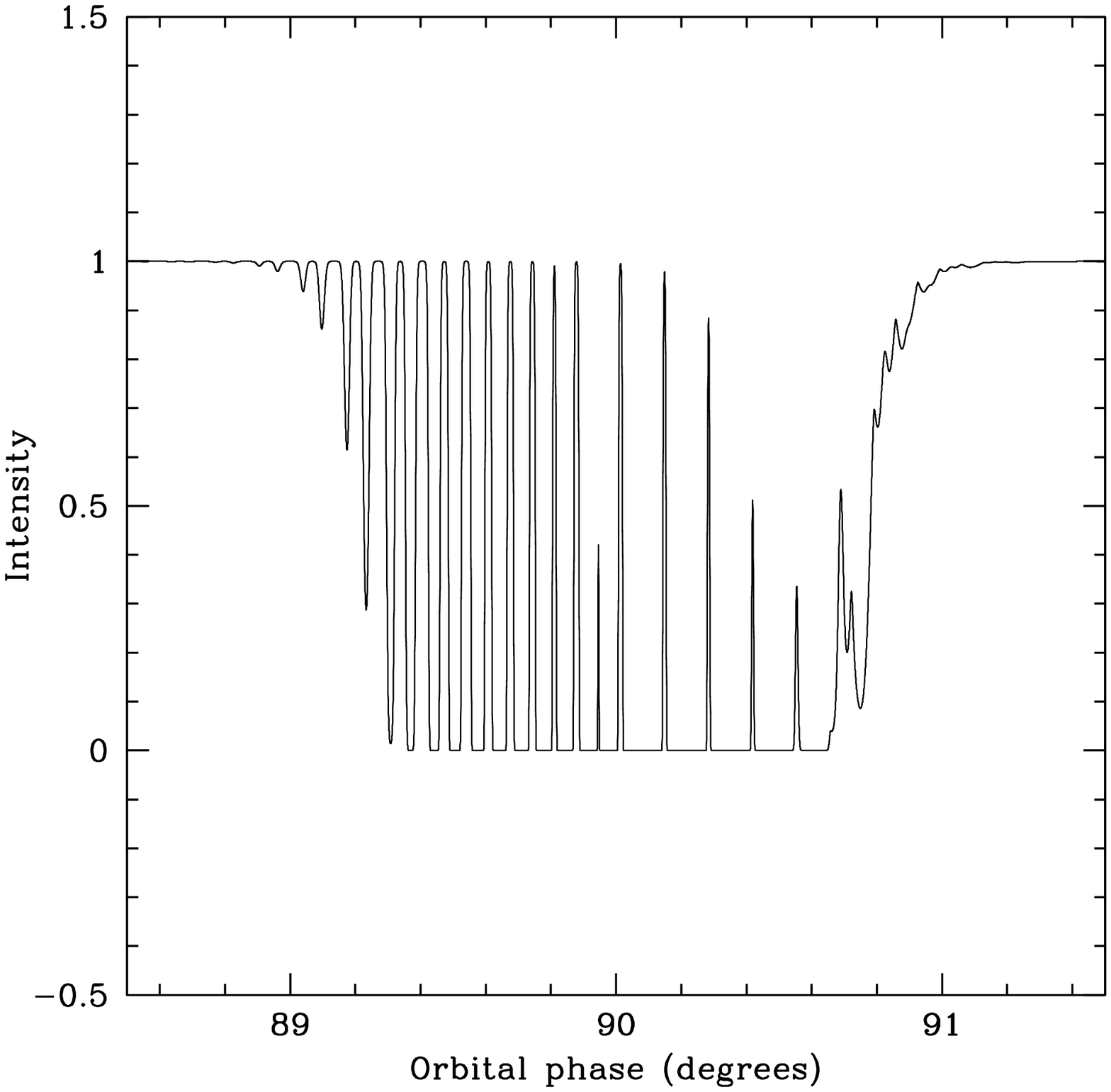}
\caption{(a) Predicted eclipse profile at high temporal resolution of 
$\Delta t = 0.01$ s at 800 MHz.  Parameters as in Fig. \ref{compare}.
The absorbing plasma has a sharp cutoff
    on dipole field lines extending beyond 
    $R_{\rm abs+} = 1.5\times 10^9$.  
    In the transparent windows, the flux returns to the unabsorbed level.
    (b) Eclipse model with electron density given by 
    eq. (\ref{neion}).  Parameters as in Fig. \ref{compareb}.
    In some transparent windows, the flux is partially absorbed.
}
\label{full}
\end{figure}

\begin{thebibliography}{}
\bibitem[Arons et al. (2004)]{arons}
Arons, J., Backer, D.~C., Spitkovsky, A., \& Kaspi, V.~M. 2004,
  astro-ph/0404159

\bibitem[Burgay et al. (2003)]{burgay03}
Burgay, M., D'Amico, N., Possenti, A., et al. 2003, \nat, 426, 531

\bibitem[Cole et al. (2004)]{coles}
Coles, W.~A., McLaughlin, M.~A., Rickett, B.~J.,
 	Lyne, A.~G., \& Bhat, N.~D.~R. 2004, astro-ph/0409204

\bibitem[Demorest et al. (2004)]{demo04}
Demorest, P., Ramachandran, R., Backer, D.~C.,
Ransom, S.~M., Kaspi, V., Arons, J., Spitkovsky, A. 2004, \apjl,
  615, 137

\bibitem[Fung \& Kuijpers(2004)]{fung04} Fung, P.~K., \&
Kuijpers, J.\ 2004, \aap, 422, 817

\bibitem[Goldreich \& Julian(1969)]{goldreich69} Goldreich, P., \&
Julian, W.~H.\ 1969, \apj, 157, 869


\bibitem[Harding et al. (2002)]{harding}
Harding, A.~K., Muslimov, A.~G., \& Zhang, B. 2002, \apj, 576, 366

\bibitem[Goldreich \& Sridhar(1997)]{goldsh97}
Goldreich, P. \& Sridhar, S. 1997, \apj, 485, 680

\bibitem[Hibschman \& Arons(2001)]{hibs01}
Hibschman, J.~A. \&  Arons, J. 2001, \apj, 560, 871

\bibitem[Holloway(1977)]{Holloway77}
Holloway, N.~J. 1977, \mnras, 181, 9


\bibitem[Kabin et al. (2004)]{kabin04}
Kabin, K. et al., 2004, Journal of Geophysical Research (Space Physics),
A18, 5222

\bibitem[Kaspi et al. (2004)]{kaspi}
Kaspi, V.~M., Ransom, S.~M., Backer, D.~C., Ramachandran, R.,
Demorest, P., Arons, J., \& Spitkovskty, A. 2004, \apj, 613, 137


\bibitem[Lyne et al. (2004)]{lyne} Lyne, A.~G., et al. 2004,
Science,  303, 1153

\bibitem[Lyutikov(2004)]{lyut}
Lyutikov, M. 2004, \mnras, 353, 1095

\bibitem[Lyutikov  et al. (1998)]{lbm98}
{{Lyutikov}, M. and {Blandford}, R.~D. and {Machabeli}, G.},
1999, {\mnras}, 305,
338

\bibitem[McLaughlin et al. (2004)]{mcla04}
McLaughlin, M.~A., et al. 2004, \apjl, 616, 131

\bibitem[Michel(1969)]{mich69}
Michel, F.~C., \apj, 158, 727

\bibitem[Pacholczyk \& Swihart(1970)]{Pachol70}
Pacholczyk, A.~G. \& Swihart, T.~L. 1970, \apj, 161, 415
Fransisco

\bibitem[Rafikov \& Goldreich(2004)]{rafikov}
Rafikov R.~R. \& Goldreich, P. 2004, astro-ph/0412355

\bibitem[Ransom et al. (2004)]{ransom04}
Ransom, S.~M., Kaspi, V.~M., Ramachandran, R.,
Demorest, P., Backer, D.~C., Pfahl, E.~D., Ghigo, F.~D.,
\& Kaplan, D.~L. 2004, \apjl, 609, 71


\bibitem[Rybicki \& Lightman(1979)]{rybiki}
Rybicki, G. \& Lightman,  A.~P. 1979,
\textit{Radiative Processes in Astrophysics},
Wiley, New York

\bibitem[Sagiv et al. (2004)]{sagiv04}
Sagiv, A., Waxman, E., \& Loeb, A. 2004, \apj, 615, 366

\bibitem[Taylor (1974)]{taylor74}
Taylor, J.~B. 1974, \prl, 33, 1139

\bibitem[Thompson et al. (1994)]{thomp94}
Thompson, C., Blandford, R.~D., Evans, C.~R., \& Phinney, E.~S.
1994, \apj, 422, 304

\bibitem[Thompson \& Blaes(1998)]{thomp98}
Thompson, C. \& Blaes, O. 1998, \prd, 57, 3219

\bibitem[Thompson et al. (2002)]{tlk02}
Thompson, C.,  Lyutikov, M., Kulkarni, S.~R. 2002, \apj, 574, 332


\end{thebibliography}
\end{document}